\documentclass[12pt, final]{article}

\usepackage[utf8x]{inputenc}
\usepackage[T1]{fontenc}

\fontsize{12pt}{18pt}\selectfont

\usepackage{amsmath,amsfonts,amssymb,amsthm}

\usepackage{bbm}
\usepackage{graphicx}
\usepackage[notref, notcite]{showkeys}
\usepackage{mathtools}
\usepackage{tabularx}
\usepackage[hyperfootnotes=false]{hyperref}
\usepackage{color}
\usepackage[a4paper, textheight=21cm, textwidth=16cm]{geometry}

\usepackage{xargs}                      
\usepackage[pdftex,dvipsnames]{xcolor}
\usepackage[colorinlistoftodos,prependcaption,textsize=footnotesize]{todonotes}

\newtheorem{thm}{Theorem}[section]
 
 \newtheorem{lemma}[thm]{Lemma}
 \newtheorem{var}[thm]{Criterion}
 \newtheorem{prop}[thm]{Proposition}

\DeclareSymbolFont{txlargeoperatorsA}{U}{txexa}{m}{n}
\DeclareMathSymbol{\fint}{\mathop}{txlargeoperatorsA}{62}

\newcommand{\uu}{\mathbf{u}}
\newcommand{\xx}{\mathbf{x}}
\newcommand{\xp}{\mathbf{x}'}

\newcommand{\vv}{\mathbf{v}}

\newcommand{\Ra}{{\rm{Ra}}}
\newcommand{\Gr}{{\rm{Gr}}}
\newcommand{\Pra}{{\rm{Pr}}}
\newcommand{\Nu}{{\rm{Nu}}}
\newcommand{\LL}{\rm{\mathcal{L}_s}}
\newcommand{\ux}{u^x}

\newcommand{\uz}{u^z}
\newcommand{\up}{\mathbf{u}'}
\newcommand{\f}{\mathbf{f}}
\newcommand{\fx}{f^x}
\newcommand{\fy}{f^y}
\newcommand{\fz}{f^z}
\newcommand{\la}{\langle}
\newcommand{\ra}{\rangle}
\newcommand{\R}{\mathbb{R}}
\newcommand{\lla}{\left\langle}
\newcommand{\rra}{\right\rangle}
\newcommand{\Q}{\mathcal{Q}^{\tau}}

\newcommand{\hatt}{\hat{\theta}}
\newcommand{\hatuz}{\hat{\uz}}
\newcommand{\Real}{\mathcal{R}}
\newcommand{\Nubf}{{\rm{Nu_{bf}}}}
\newcommand{\DeltaH}{\Delta_{{\rm H}}}

\pagecolor{white}

\newcommand\blfootnote[1]{%
  \begingroup
  \renewcommand\thefootnote{}\footnote{#1}%
  \addtocounter{footnote}{-1}%
  \endgroup
}

\numberwithin{equation}{section}

\makeatletter
\renewcommand{\@biblabel}[1]{#1\hfill \hspace{-0.2cm}}
\makeatother

\begin{document}

\phantom{ }
\vspace{4em}

\begin{flushleft}
{\large \bf The role of boundary conditions in scaling laws for turbulent heat transport}\\[2em]
{\normalsize \bf Camilla Nobili$^1$}\blfootnote{\today}\\[0.5em]
\small \begin{tabbing} 
$^1$ \= Department of Mathematics, University of Hamburg, 20146 Hamburg, Germany.\\
\>E-mail: camilla.nobili@uni-hamburg.de\\
\end{tabbing}
\end{flushleft}

{\small
{\bf Abstract:}
In most results concerning bounds on the heat transport in the Rayleigh-B\'enard convection problem no-slip boundary conditions for the velocity field are assumed. Nevertheless it is debatable, whether these boundary conditions reflect the behavior of the fluid at the boundary. This problem is important in theoretical fluid mechanics as well as in industrial applications, as the choice of boundary conditions has effects in the description of the boundary layers and its properties. In this review we want to explore the relation between boundary conditions and heat transport properties in turbulent convection. 
For this purpose, we present a selection of contributions in the theory of rigorous bounds on the Nusselt number, distinguishing and comparing results for no-slip, free-slip and Navier-slip boundary conditions.  
}


\vspace{2em}

\begin{center}
\textit{ {In memory of Charlie Doering and his seminal works on turbulent convection}}
\end{center}
\section{Introduction}
The Rayleigh-B\'enard convection model describes a buoyancy driven flow of a fluid heated from below and cooled from above. In a finite box with height $h$, the impermeable horizontal bottom an top plates are held at temperature $T_b$ and $T_t$ respectively, with $T_t<T_b$. Temperature differences trigger density differences within the fluid layers, which, in turn, generate convective motions. As the intensity of the buoyancy forces grows, the dynamics undergo a series of bifurcations: convection rolls are destabilized by plume-shaped structures forming at the boundaries and the motion becomes unpredictable, chaotic, turbulent. This model is a paradigm of nonlinear dynamics and pattern formation and has a myriad of applications in engineering of heat transfer \cite{D20-b}. The motion within the finite box is governed by an advection-diffusion equation for the temperature field $T$ coupled with the Navier-Stokes equations for the velocity field $\uu$. Unless otherwise stated, we will focus on three-dimensional dynamics and denote by $\xx=(\xp,z)=(x,y,z)$ a vector in $\R^3$ and by $\mathbf{e_z}=(0,0,1)^t$ the vertical normal vector.  In dimensionless variables \footnote{see the Boussinesq equations in \cite[Section 2.1]{DG} and their non-dimensionalization choosing length scales determined by the vertical gap height $h$, time scales defined by the thermal diffusion $\kappa$ and temperature scales determined by the temperature gap $T_b-T_t$.}, the velocity field $\uu=(\up, u^z)=(u^x,u^y,u^z):\R^3\times (0,\infty)\rightarrow \R^3 $ and the scalar temperature $T:\R^3\times(0,\infty)\rightarrow \R$ evolve
in the domain $\Omega=[0,L_x]\times [0,L_y]\times [0,1]$ according to 

\begin{equation}\label{RBC}
\begin{array}{rll}
  \partial_{t} T+\uu\cdot \nabla T-\Delta T &=&0\\
  \nabla \cdot \uu&=&0\\
  \frac{1}{\Pra}(\partial_{ t} \uu+\uu\cdot \nabla \uu)+\nabla p-\Delta\uu&=&\Ra T  \mathbf{e_z}\,.\\
\end{array}
\end{equation}
This is also called \textit{Boussinesq system} as, in its derivation, density variations generating the vertical buoyancy force appearing in the right-hand side of the velocity equation, are neglected elsewhere in the velocity and temperature equation (Boussinesq approximation).
Two dimensionless parameters appear in the system after the non-dimensionalization:
\begin{equation*}
  \Ra:=\frac{\alpha g(T_b-T_t)h^3}{\kappa\nu}
\end{equation*}
 is the (control parameter) Rayleigh number and 
\begin{equation*}
  \Pra:=\frac{\nu}{\kappa}
\end{equation*}
is the (material parameter) Prandtl number. Here $\alpha$ is the thermal expansion coefficient, $g$ is the acceleration of gravity, $\nu$ is the kinematic viscosity and $\kappa$ is the thermal diffusivity. The pressure field $p(\xx,t)$ appears as a Lagrange multiplier enforcing the incompressibility condition $(\nabla \cdot \uu=0)$. 
In the non-dimensional varibles, the temperature boundary conditions are
\begin{equation}\label{Tbc}
T=1 \;\mbox{ at }\; z=0 \quad \mbox{ and } \quad T=0 \;\mbox{ at }\; z=1\,. 
\end{equation}
For technical convenience we will suppose all functions to be periodic in the horizontal variables $\xp$ with period $L=(L_x,L_y)$.
We refer to \cite{OPN15} for details about the derivation of the system.

One of the most important open problems in fluid dynamics is to determine the rate at which heat is transferred in turbulent convection. This theoretical problem finds uncountable many applications in meteorology, oceanography and industry \cite{D20-b}.
 The natural nondimensional  quantity for gauging heat transfer in the upward direction is the Nusselt number $\Nu$ defined as the ratio between the convective to the conductive heat flux
\begin{equation}\label{Nu-def1}
  \Nu=\frac{\lla\int_0^1 (\uu T-\nabla T)\cdot \mathbf{e_z}\, dz\rra}{\lla\int_0^1 -\nabla T\cdot \mathbf{e_z}\, dz \rra}=\lla\int_0^1 (\uu T-\nabla T)\cdot \mathbf{e_z}\, dz\rra\,,
\end{equation}
where

$$\la f\ra=\limsup_{t\rightarrow \infty}\frac{1}{t}\int_0^t\frac{1}{L_xL_y}\int_0^{L_x}\int_0^{L_y}f(\cdot, z)\, dx\, dy\, dt\,.$$
By similarity law, the Nusselt number $\Nu$ must be a function of the non-dimensional parameters appearing in the system, and thus obeys a functional relation of the type
$$\Nu=f(\Ra,\Pra)\,.$$
Theoretical predictions (Malkus \cite{MA54}, Kraichnan \cite{Kra62}, Spiegel \cite{Spi71}, Siggia \cite{S94}) and experiments (see \cite{AGL09} and references therein) suggest a power-law scaling of the form
$$\Nu \sim \Ra^{\alpha} \Pra^{\beta} \qquad \mbox{ for } \alpha, \beta \in \R\,,$$
where $\alpha$ and $\beta$ depend on the characteristic of the flow. 
In Section \ref{prelim} we will illustrate two physically motivated heuristic arguments predicting different scaling behaviors. The challenge for physicists, mathematicians, and engineers is to identify the value of the powers $\alpha$ and $\beta$, as $\Ra$ and $\Pra$ varies.
 As described by the stability theory, there exists a critical Rayleigh number, $\Ra_c$, under which the pure conduction profile is stable. As Rayleigh number grows over this critical value,  convection rolls appears and are eventually destabilized by plumes-like structures arsing from the boundary layers. At very high Rayleigh numbers the dynamics become turbulent. These three phases can be appreciated in Figure 1. As heat transport is enhanced in turbulent regimes, in the whole manuscript we will assume $\Ra\gg 1$. The Prandtl number, instead will be mostly assumed to be large, because of limitation in our analysis. In fact, the smaller the Prandtl number, the stronger is the inertial term in the Navier-Stokes equations.

In this paper we will review some of the major contributions 
concerning bounds for the Nusselt number in the Rayleigh-B\'enard convection problem \eqref{RBC}--\eqref{Tbc} in the last thirty years, distinguishing between no-slip, free-slip and Navier-slip type boundary conditions for the velocity field.
Most numerical simulations and rigorous results assume no-slip boundary conditions in the horizontal plates and periodic sidewalls. Whether or not the no-slip conditions are the more suitable for this problem, remains a matter of debate. This is a general and very important question that has been posed by the founders of fluid dynamics and is still subject of ongoing research. Regarding the issue of detecting suitable boundary conditions for the Navier-Stokes equations, Serrin \cite{Serrin} writes ``\textit{The conditions to be satisfied at a solid boundary are more difficult to state, and subject to some controversy. Stokes argued that the fluid must adhere
to the solid, since the contrary assumption implies an infinitely greater resistance to the sliding of one portion of fluid past another than to the sliding of fluid
over a solid. Another and perhaps stronger argument in favor of adherence,
at least for the case of liquids at ordinary conditions, is found in experiments with
tube viscometers where the fourth power of the diameter law is conclusively
verified. Although these facts are quite convincing when moderate pressures
and low surface stresses are involved, they do not apply in all cases; indeed in
high altitude aerodynamics an adherence condition is no longer true}''. Many works took inspiration from Serrin's observation. Here we mention \cite{APS13}, where the authors addressed theoretical questions related to the choice of boundary conditions.

\begin{figure}[h!]
\centering
\includegraphics[scale=0.8]{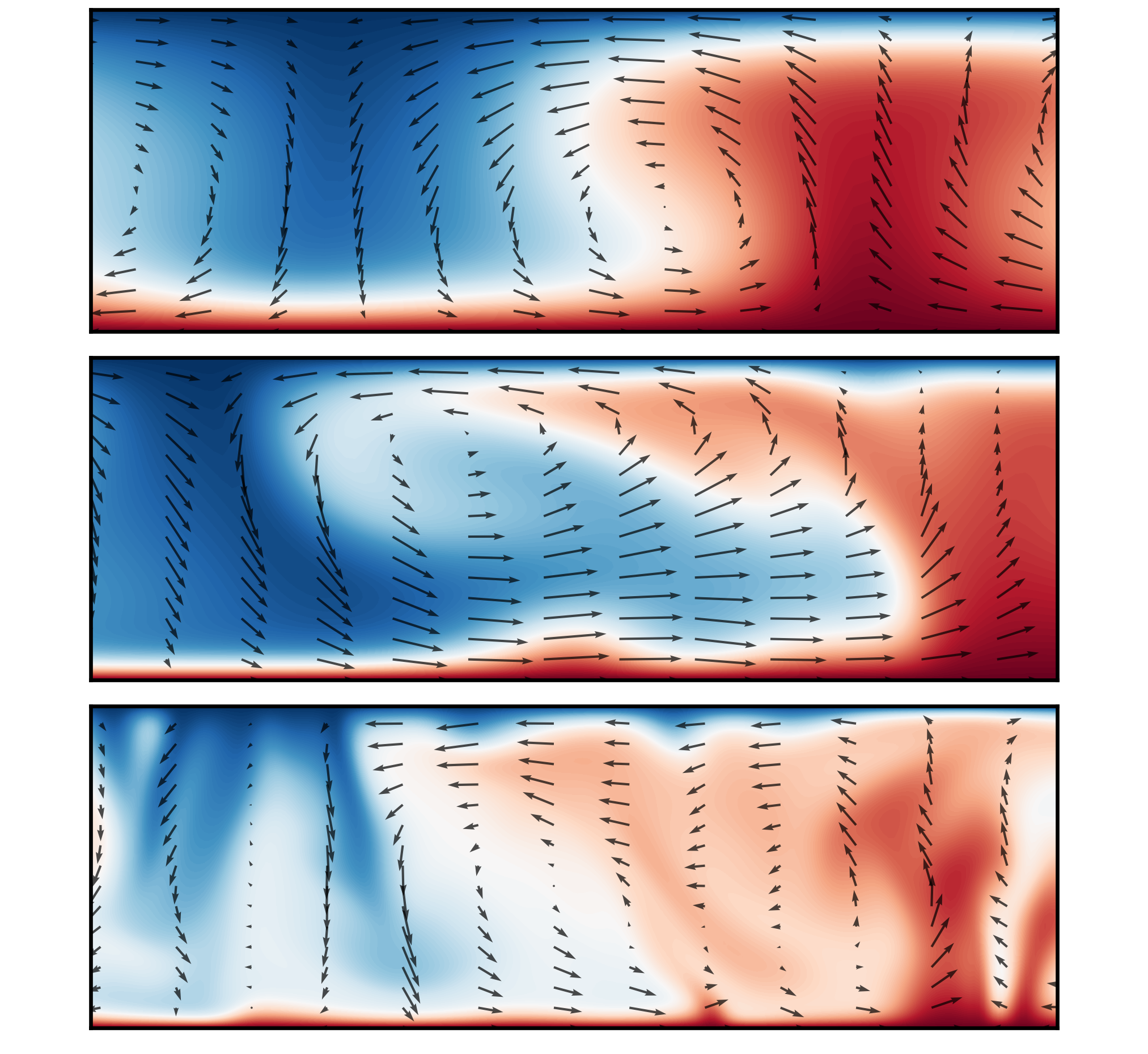}
\caption{Three vertical slices in the $xz$-plane of three-dimensional, free-slip, simulation in \cite{VSS21} at $\Ra = 3850$, $\Ra=38501$, and $\Ra=385014$ and $\Pr = 1$. Each panel contains the instantaneous temperature field in the background with colors from blue (cold) to red (hot).
The arrows project the instantaneous velocity onto the 2d-plane. The pictures describe three states as $\Ra$ increases: convection rolls, destabilization caused by plumes arising from the boundary layers and turbulent convection.}
\end{figure}

In the last decade, free-slip boundary conditions for the Rayleigh-B\'enard convection problem have been considered and rigorous upper bounds on the Nusselt number indicate the possibility of enhanced heat transport compared to the no-slip case.  But which one of those boundary conditions better represent the physical picture?
On the one hand, as remarked in \cite{WD12}, (especially) in ``turbulent regimes'' the boundary conditions might not be perfectly no-slip. On the other hand, the mathematically interesting free-slip boundary conditions are non-physical because of the absence of vorticity production at the boundary, as we will see later in Section \ref{four}. H. Navier in 1824  \cite{N1827} suggested the boundary conditions
\begin{equation}\label{NS-cond}
2\nu [D\uu \cdot \bar n]\cdot \bar t + \alpha \uu\cdot \bar t = 0  
\end{equation}
where $D$ is the rate of strain tensor, $\bar t$ and $\bar n$ are the tangential and normal vector to the boundary and $\alpha>0$ is the friction coefficient. These conditions go by the name of \textit{Navier-slip boundary conditions} and can be interpreted as an interpolation between no-slip and free-slip boundary conditions. Since, as observed in \cite{WW13}, very few surfaces are truly no-slip or free-slip, the Navier-slip boundary conditions appear reasonable and have been studied in a multitude of applied and theoretical problems, among which we want to mention the works on vanishing viscosity limits for the two-dimensional Navier-Stokes equations \cite{CMR98,FLP05}.

In this manuscript we will focus on
\begin{enumerate}
  \item  No-slip boundary conditions: 
  \begin{equation}\label{no-slip}\uz=0\,, \qquad \up=0 \quad \mbox{ at } z=\{0,1\}\,.\end{equation}
  \item  Free-slip boundary conditions: 
  \begin{equation}\label{free-slip}\uz=0\, \qquad \partial_z \up=0  \quad \mbox{ at } z=\{0,1\}\,.\end{equation}
  \item  Navier-slip boundary conditions: 
  \begin{eqnarray}\label{Navier-slip}
  &&\uz=0 \quad \mbox{ at } z=\{0,1\}\,,\notag\\
  &&\partial_z \up=-\frac{1}{\LL}\up \quad\mbox{ at } z=1 \quad \mbox{ and } \qquad  \partial_z \up=\frac{1}{\LL}\up \quad \mbox{ at } z=0\,,
  \end{eqnarray}
  where $\LL$ is the slip length.
\end{enumerate}
For these three types of boundary conditions we will critically review results concerning rigorous bounds on the Nusselt number and highlight differences in the analyses.
As we will see later, while the model equipped with no-slip boundary conditions for $\uu$ is most commonly investigated, the other two types of boundary conditions and their role have been less explored. The goal of this review is to show how the change of boundary conditions affects heat transport properties of the fluid. 

\medskip

The review is organized as follows: In Section 2 we will go through two significant physical arguments predicting different scaling behavior for the Nusselt number. Moreover, starting from  definition \eqref{Nu-def1} and using a-priori estimates, we will obtain other identities involving the Nusselt number.  Depending from the strategy adopted to produce bounds on the Nusselt number (introduced in the third subsection), each identity will later be used singularly or combined.  Section 3 is divided in two parts: in Subsection 3.1 we will review some important contributions for the infinite Prandtl number system. In particular we will introduce the background field method and discuss its success and limitation. Then we will present an alternative argument proposed by Constantin and Doering in 1999 and show how this led Otto and Seis in 2011 to produce the optimal (so far) upper bound on the Nusselt number for no-slip Rayleigh-B\'enard convection at infinite Prandtl number. In Subsection 3.2 we will go through three arguments to derive bounds for the Nusselt number in the finite Prandtl number case. 
Rigorous results for the system complemented with the other types of boundary conditions considered in this paper, namely free-slip and Navier-slip boundary conditions, are reviewed in Section 4.   
In all sections we will examine results in a (partial) chronological order and try to guide the reader through the reasoning behind development of new strategies and, eventually, optimal results. 

\section{Preliminaries}\label{prelim}

\subsection{Physical scalings}
In the fifties Malkus \cite{MA54}, considering fluids with very high viscosity,
performed experiments in which he noticed sharp transitions in the slope of the $\Nu-\Ra$ relation and suggested  the scaling 
\begin{equation}\label{Ra13}
  \Nu\sim \Ra^{\frac 13}
\end{equation}
for very high Rayleigh numbers by a \textit{marginal stability argument}. This is based on the assumption that the heat flux is limited by transport across the (emergent) thermal boundary layer.
We illustrate Malkus' argument here.
Since the main temperature drop happens near the boundary, we can assume that in a bottom boundary layer of thickness $\delta$ (to be determined)
the temperature drops from $1$ to its average $\frac 12$. Thanks to the average $\langle\cdot\rangle$, we may extract 
the Nusselt number from the boundary layer, where by the no-slip boundary condition we have $(\uu T-\nabla T)e_z\approx \partial_z T$
so that $\Nu\sim \frac{1}{\delta}$. 
We think of the boundary layer as a pure conduction state in the interval $0\leq z\leq \delta$.
Marginal stability refers to the assumption that this state is borderline stable, meaning that its Rayleigh number is critical, which in view of the definition of the Rayleigh number, means 
$$\Ra_c=\frac{g\alpha (T_b-T_t)(\delta h)^3}{\nu \kappa}\,,$$
from which, because $\Ra_c\sim 1$, we infer $\delta\sim \Ra^{-\frac 13}\,.$
Inserting this in the scaling of Nusselt number above one finds
$$\Nu\sim \Ra^{\frac 13}\,.$$
The same conclusion can be achieved by rescaling the equations according to
\begin{equation}\label{rescaling1}
  \xx=\Ra^{-\frac{1}{3}}\tilde \xx, \; t=\Ra^{-\frac{2}{3}}\tilde t,\; \uu=\Ra^{\frac{1}{3}}\tilde \uu, \; p=\Ra^{\frac{2}{3}}\tilde p \; \mbox{ and thus } \;\Nu=\Ra^{\frac{1}{3}}\widetilde{\Nu}\,.
\end{equation}
In this way, neglecting the inertial term in the Navier-Stokes equations, we end up with the parameter-free system  
\begin{equation*}
\begin{array}{rclc}
  \partial_{\tilde t}  T+\tilde \uu\cdot \tilde{\nabla}  T &=& \tilde{\Delta}T  \\
    -\tilde\Delta \tilde \uu+\tilde\nabla \tilde p&=& T \mathbf{e_{\tilde z}} \\
    \tilde{\nabla}\cdot \tilde \uu&=& 0 \,.\\
\end{array}
\end{equation*}
Since for the latter system, it is natural to expect that the heat flux is universal, i.\ e.\ $\widetilde{\Nu}\sim 1$, we also obtain
$\Nu\sim \Ra^{\frac{1}{3}}.$
Most recently Malkus's theory was refereed to as \textit{classical theory} \cite{D20-b}.
  
Kraichnan in \cite{Kra62} and Spiegel \cite{Spi71}, instead, proposed the scaling
\begin{equation}\label{Ra12}
  \Nu\sim \Pra^{\frac 12}\Ra^{\frac 12}
\end{equation} based on the assumption that the heat flux is limited by the transport of the fluid across the bulk. Recalling that, by definition, the convective and conductive heat flux are $q_{\rm{conv}}\sim \rho vc\Delta T$ and $q_{\rm{cond}}\sim \frac{\rho c \kappa \Delta T}{h}$ respectively and assuming
$$v\sim \sqrt{g\alpha \Delta T h} \qquad \mbox{ free fall velocity}\,, $$
then, the Nusselt number is 
$$\Nu\sim \frac{q_{\rm{conv}}}{q_{\rm{cond}}}\sim \frac{\rho \,\sqrt{g\alpha \Delta T h}\,c\Delta T}{\frac{\rho c \kappa \Delta T}{h}}=\left(\frac{g\alpha \Delta T h^3}{\kappa \nu}\right)^{\frac 12}\left(\frac{\nu}{\kappa}\right)^{\frac 12}=\Ra^{\frac 12}\Nu^{\frac12}\,.$$ 
We can obtain the same scaling by neglecting the diffusivity and the viscosity
term in the model
\begin{equation*}
\begin{cases}
  \partial_t T+\uu\cdot \nabla T=0  &
  \\
  \frac{1}{\Pra}(\partial_t \uu+(\uu \cdot \nabla)\uu)+\nabla p=\Ra T \mathbf{e_{z}} &
  \\
  \nabla\cdot\uu = 0\,, &
\end{cases}
\end{equation*}
and rescaling according to
$$t=\frac{1}{(\Pra\Ra)^{\frac 12}}\tilde{t},\; \uu=(\Pra\Ra)^{\frac 12}\tilde{\uu}, \; p=\Ra\tilde{p}\; \mbox{ and thus } \Nu=(\Pra\Ra)^{\frac 12}\widetilde{\Nu}\,.$$
Then, imitating the previous argument, the Nusselt number for the parameter-free system 
\begin{equation*}
\begin{cases}
  \partial_{\tilde{t}} T + \tilde{\uu} \cdot \nabla T = 0 &
  \\
  \frac{1}{\Pra}(\partial_{\tilde{t}} \tilde{\uu} + (\tilde{\uu} \cdot \nabla)\tilde{\uu}) + \nabla \tilde{p} = \Ra T \mathbf{e_{z}} &
  \\
  \nabla \cdot \tilde{\uu} = 0 
\end{cases}
\end{equation*}
is $$\Nu \sim \Pra^{\frac 12} \Ra^{\frac 12}\,.$$ 
The Kreichnan-Spiegel theory was recently referred to as \textit{ultimate theory} \cite{D20-b}.
\newline
In \cite{Spi71}, Spiegel writes ``\textit{What is suggested by
these arguments is that at sufficiently high $\Ra$, turbulent breakdown of the
thermal boundary layer occurs and causes a transition from \eqref{Ra13} to \eqref{Ra12}. No
such transition has been detected experimentally, but this is presumably
explained by the limitation of the experiments to what in stellar terms are
modest Rayleigh numbers. The need to confirm (or deny) \eqref{Ra12},
which is intimately connected with basic ideas of stellar structure theory,
poses a great challenge to the experimentalists.}''.
Malkus's scaling $\Nu\sim \Ra^{\frac 13}$ has been been confirmed by experiments at (relatively) 
high Prandtl numbers (cf.\ \cite{AGL09} for a list of experimental 
results) while (up to today) it remains no indication of the Spiegel-Kraichnan scaling  $\Nu\sim \Pra^{\frac12}\Ra^{\frac 12}$.

\subsection{Other representations of the Nusselt number}

The Nusselt number defined in \eqref{Nu-def1} admits other useful representations. As we will see later in the preliminaries, their employment depends on the method chosen to bound the Nusselt number. 
First, let us notice that if the initial temperature $T(\xx,0)=T_0(\xx)$ is chosen such that $0\leq T_0(\xx)\leq 1$, then, by the standard maximum principle for parabolic equations, we have 
\begin{equation}\label{MP}
  0\leq T(\xx,t)\leq 1 \qquad \mbox{ for all } t\geq 0\,.
\end{equation}
The first representation can be obtained by taking the long-time and horizontal average of the temperature equation and using $\frac 12\lla\partial_tT^2\rra=0$ as a consequence of the maximum  principle, we have 
$$
\partial_z\la \uz T- \partial_zT \ra=0\,.
$$
which implies that for all $z\in (0,1)$
\begin{equation}\label{Nu-inter}
  \la \uz T-\partial_z T \ra=\la - \partial_zT |_{z=0}\ra\,.
\end{equation}
Recalling the definition of Nusselt number in \eqref{Nu-def1}, this identity yields
\begin{equation}\label{average}
  \Nu=\la \uz T-\partial_z T \ra \qquad \mbox{ for all } z\in(0,1)\,.
\end{equation}
Therefore the Nusselt number is independent of $z$, and it is the same for each horizontal layer of fluid.
Another representation is derived by testing the temperature equation with $T$ and integrating by parts. In fact, using the incompressibility together with the temperature boundary 
conditions and \eqref{Nu-inter} we obtain
\begin{equation}\label{Nu-def2}
  \Nu=\lla\int_0^1|\nabla T|^2\, dz\rra \,.
\end{equation}
Instead, formally testing the Navier-Stokes equations with $\uu$ and integrating by parts we have 
\begin{equation}\label{velocity-balance}
  \frac 12\frac{1}{\Pr}\frac{d}{dt}\|\uu\|_{L^2}^2=-\|\nabla \uu\|_{L^2}^2+\Ra\int_{\Omega} T \uz\,.
\end{equation}
Note that this energy balance (formally) holds for all boundary conditions considered in this article (i.e. \eqref{no-slip},\eqref{free-slip},\eqref{Navier-slip}) as it uses only that $\uz=0$ at $z=\{0,1\}$ and the incompressibility condition.
\newline
If the energy of the velocity remains bounded in time, averaging in \eqref{velocity-balance} we obtain 
\begin{equation}\label{diss-bound}
  \lla\int_0^1|\nabla \uu|^{2}\, dz\rra=\Ra\lla\int_0^1 T\uz\, dz\rra=\Ra(\Nu-1)\,.
\end{equation} 
This identity produces yet another useful representation of the Nusselt number
$$\Nu=1+\frac{1}{\Ra}\lla\int_0^1|\nabla \uu|^2\, dz\rra\,.$$
Clearly this computation is rigorous in two dimensions, but only formal in three dimensions. In this last case, defining Leray-solutions we will obtain an energy \textit{inequality} in \eqref{diss-bound}.

\subsection{Methods to bound the Nusselt number}
In this section we briefly introduce the reader to the most common techniques used to bound the Nusselt number. Some of these methods will be carefully explained and used in the next sections.

The first rigorous result was proven by Howards \cite{H1963} by transforming the problem of finding upper bounds on $\Nu$ into a variational problem. Howard found
$$\Nu\lesssim \Ra^{\frac 12}$$ under the hypothesis of ``statistical stationarity'', meaning that horizontal averages are time-independent \footnote{this holds for stationary flows}. Later Busse extended Howard's result to solutions with multiple boundary-layer structure \cite{B1969} obtaining the same bound with an improved constant prefactor.

The \textit{background field method}, proposed (in this context) by Doering and Constantin in \cite{DC96}, consists in decomposing the temperature field $T$ as 
\begin{equation}\label{dec}
T(\xp,z,t)=\tau(z)+\theta(\xp,z,t)\,,
\end{equation}
where  $\tau$ is a steady \textit{background profile} satisfying the driven boundary conditions
$$\tau=1 \quad \mbox{ at }\; z=0 \qquad \mbox{ and }\qquad \tau=0 \quad \mbox{ at }\; z=1 $$
and $\theta$ represents the temperature fluctuations satisfying
\begin{equation}\label{temp-fluc-bc}
\theta=0 \qquad \mbox{ at }\; z\in\{0,1\}\,.
\end{equation}
The fluctuations evolve according to
\begin{equation}\label{flactuation-equation}
 \partial_t \theta+\uu\cdot\nabla\theta=-\tau'u^z+\Delta\theta+\tau''
\end{equation}
and the using the representation \eqref{Nu-def2}, the Nusselt number can be rewritten as
\begin{equation}\label{temp-Nu}
 \Nu=\int_0^1|\tau'|^2\, dz-\Q\{\theta,\uu\}\,,
\end{equation}
where, for fixed $\tau$, $\Q$ is a quadratic functional in $\theta$ and $\uu$.
In this way we can transform the problem of finding the optimal upper bound on the Nusselt number into a variational problem: 
find the background profile $\tau$ for which $\Q\{\theta,\uu\}\geq 0$ for all $\theta$ and minimizes the integral $\int_0^1|\tau'|^2\, dz$.
The background field method will be carefully described in Section \ref{total} and we 
 refer to \cite{FAW21} for a recent review on it.
\newline
A generalization of the background field method was proposed in \cite{CGHP14} to improve the available bounds. The method is based on the following observation: if $V(t)=V\{\theta(\cdot, t), \uu(\cdot, t)\}$ is a uniformly in time bounded and differentiable function with $\lim_{t\rightarrow \infty}\int_0^t\frac{d}{ds}V(s)\, ds=0$ 
then, in order to prove 
\begin{equation}\label{aux}
\Nu= \lla\int_0^1|\nabla T|^2\, dz\rra\leq B\,,
\end{equation}
it suffices to show 
\begin{equation*}
\frac{d}{dt}V+\lla\int_0^1|\nabla T|^2\, dz\rra_{\rm{H}}-B\leq 0\,,
\end{equation*}
where 
$$\la f(\cdot,z)\ra_{\rm{H}}=\frac{1}{A}\int_0^{L_x}\int_0^{L_y}f(\cdot, z)\, dx\, dy\,.$$

In fact, time-averaging this inequality we recover \eqref{aux}. It is easy to see that the auxiliary functional method and the background field method coincide when the auxiliary functional $V$ is quadratic in $\uu$ and $\theta$ \cite{C17}. The power of this method is to allow the functional $V$ to be more general (than the quadratic form $\Q$), to the expenses of an increasing analytical complexity which eventually requires computer-assisted investigation \cite{G18,FGHC16}.  The auxiliary functional method has proven to be successful to produce sharp bounds for (system) of ordinary differential equations \cite{TGD17}, but the application of this method for producing optimal bounds on the Nusselt number in the Rayleigh-B\'enard convection can only be speculated.  

Another method to produce upper bounds based on the vertical localization of the Nusselt number was
first used in \cite{OS11} and then proposed by \cite{S15} as an alternative method to the background field method for a variety of fluid dynamics problems.  This method is based on the observation that \eqref{average} can be averaged over \textit{any} horizontal layer of fluid; in particular, 
averaging near the bottom boundary layer $0\leq z\leq\delta$ we are led to the definition
\begin{equation*}
  \Nu=\frac{1}{\delta}\int_0^{\delta} \langle u^zT-\partial_z T\rangle\, dz\,.
\end{equation*} 
\footnote{Notice here that, if $|\int_0^t\iint u^zT\, d\xp\, ds|<g(z)\in L^1$, then the switch in order of the $\lla\cdot\rra$ averages and the integral in $z$ is justified by the Lebesgue dominated convergence theorem. The condition will be generally satisfied thanks to the a-priori (energy) estimates for $u$ and $T$.} 
In particular, because of the boundary conditions on the temperature and the fact that $T\geq 0$ for every $\xx, t$, by the maximum principle,
we have
\begin{equation}\label{av-Nusselt}
  \Nu\leq\frac{1}{\delta}\int_0^{\delta} \langle u^zT\rangle\, dz+\frac{1}{\delta}\,.
\end{equation}
The goal is now to exploit the equations of motion to derive regularity bounds on $\uz$ and $T$ in the strip $[0,L_x]\times[0,L_y]\times[0,\delta]$. As we will see later in Section \ref{FPNs}, identity \eqref{diss-bound} will be particularly useful, when adopting this method. Compared to the background field method, this approach has the advantage to be free from the rigid constraint imposed by the variational method. Nevertheless in Section \ref{total} we explain that these two methods coincide for a particular choice of background profile.


\section{No-slip boundary conditions}
\subsection{Infinite Prandtl number}\label{ipns}
For some fluids the Prandtl number is so high (see \cite{S94}, table 1) that the effect of inertia in the Navier-Stokes equations is almost negligible.  This motivates the study of the infinite-Prandtl number model, obtained by setting $\Pra=\infty$ in the momentum equation in \eqref{RBC}
 \begin{eqnarray}\label{Stokes}
 -\Delta \uu+\nabla p&=&\Ra T e_z\\
\uu&=&0\notag\,.
 \end{eqnarray}
 Using the incompressibility condition it is easy to see that \eqref{Stokes} can be rewritten as the fourth-order elliptic equation
 \begin{eqnarray}\label{4order-elliptic}
 -\Delta^2 \uz&=&\Ra \Delta_{{\rm H}}\theta\\
 \uz=\partial_z\uz&=&0\,,\notag
 \end{eqnarray}
 where $\DeltaH=\partial_x^2+\partial_y^2\,$ and the pressure is eliminated. Notice that by the incompressibility condition we infer that $\partial_z\uz=-\nabla_H\cdot \up=0$ at $z=\{0,1\}$ since $\up=0$ at $z=\{0,1\}$.
 From relation \eqref{4order-elliptic}, it is now clear that the temperature is ``instantaneously'' slaved to the velocity field.
Xiaoming Wang in \cite{WA04,WA07} reasonably asked whether this reduction is justified and whether the statistical properties of the flow in the $\Pra=\infty$ model and in the finite (but large) $\Pra$ number model can be close in some sense. The author proved that, indeed, the global attractor of the Boussinesq system at large Prandtl number converges to that of the infinite-Prandtl number model, giving some indication on the possible closedness of the respective statistical properties.

\subsubsection{Story of the background field method in five acts}\label{total}
\paragraph{Introduction of the background field method:}
The first application of the background field method to find upper bounds on the Nusselt number for the Rayleigh-B\'enard convection problem appears in the Doering and Constantin 1996 paper \cite{DC96}. We reproduce their arguments, adapting them to the infinite Prandtl number case.
We start by splitting $T$ according to \eqref{dec}.
Testing equation (\ref{flactuation-equation}) with $\theta$ and integrating by parts in $\Omega$ we obtain
\begin{equation}\label{pert-balance}
\frac 12\frac{d}{dt}\|\theta\|_{L^2}^2=
-\int_{\Omega}\tau'u^z\theta-\|\nabla\theta\|_{L^2}^2+\int_{\Omega}\tau''\theta\,,
\end{equation}
where we used incompressibility together with the condition $\uz=0$ at $z=\{0,1\}$ to eliminate the term  $\frac{1}{2}\int_{\Omega}u\cdot\nabla\theta^2$;
then, averaging we get
\begin{equation*}
\lla\int_0^1\tau'\partial_z\theta\, dz\rra=-\lla\int_0^1|\nabla\theta|^2\, dz\rra-\lla\int_0^1\tau'\theta \uz\, dz\rra\,.
\end{equation*}
The combination of this identity with 
\begin{equation}\label{r-dec}
\Nu\stackrel{\eqref{Nu-def2}}{=}\lla \int_0^1|\nabla T|^2\, dz\rra=\int_0^1|\tau'|^2\, dz+2\lla\int_0^1\tau'\partial_z\theta\, dz\rra+\lla \int_0^1|\nabla \theta|^2\, dz\rra\,,
\end{equation}
yields
\begin{equation}\label{Nu-rep-bfm}
\Nu=\int_0^1|\tau'|^2\, dz-\lla\int_0^1|\nabla\theta|^2\, dz\rra-2\lla\int_0^1\tau'\uz\theta\, dz\rra\,.
\end{equation}
Thanks to this new representation of the Nusselt number we can formulate the following
\begin{var}
 If $\tau:[0,1]\rightarrow\R$ satisfies
\begin{equation}\label{spectral-const}
  \Q\{\theta\}=\lla\int_0^1|\nabla\theta|^2\, dz+2\int_0^1\tau'\uz\theta\, dz\rra\geq 0\,,
\end{equation}
for all fluctuations $\theta(\xp, z)$ satisfying \eqref{temp-fluc-bc} and related to $\uz$ through \eqref{4order-elliptic}, then
\begin{equation}\label{Denergy}
  \Nu\leq \int_0^1|\tau'|^2\, dz\,.
\end{equation}
\end{var}
Condition \eqref{spectral-const} is often referred to as ``stability constraint''  or ``spectral constraint'' of the background profile.
The strategy just described goes by the name of \textit{background field} (or \textit{background flow}) method. Already introduced in 1992 by Constantin and Doering to bound the energy dissipation in shear driven turbulence \cite{DC92}, this method has been extensively used for estimating the Nusselt number. In fact, an upper bound is immediately obtained by computing the Dirichlet energy of a stable (in the sense specified above) background profile. Thus, each profile satisfying \eqref{spectral-const}, produces an upper bound for the Nusselt number. 
\newline
The first example of a \textit{marginally} stable profile is (a smooth approximation of)
\begin{equation}\label{choice0}
  \tau(z)=
  \begin{cases}
  1-\frac{z}{\delta} & \quad 0\leq z\leq \delta\\
  0 & \quad \delta\leq z\leq 1\,,\\
\end{cases}
\end{equation}
for which one easily shows $\Nu\lesssim \Ra^{\frac 12}$.
\begin{figure}[h!]
\centering
\includegraphics[scale=0.4]{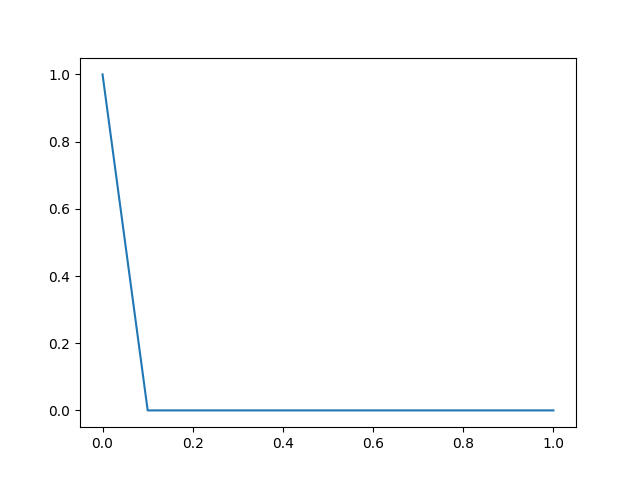}
\caption{Example of background profile $\tau=\tau(z)$ marginally stable for $\delta\sim \Ra^{-\frac 12}$.}
\end{figure}
 We use this example to explain the adjective ``marginal'' in the stability condition. In analogy to the Malkus' argument reported in Section \ref{prelim}, this refers to the fact that the boundary layer $[0,\delta]$ adjusts itself to be ``borderline'' stable. We are going to see that the thickness of $\delta$ determines the validity of \eqref{spectral-const} and eventually the bound on the Nusselt number. The bigger $\delta$ is, the smaller the upper bound on the Nusselt number. Obviously there will be artificially thin boundary layers that will ensure the stability condition, but for optimality of the upper bound, we want to detect the thickest stable boundary layer. Any boundary layers thicker than this, will be unstable.

Now we illustrate the idea behind the proof of the marginal stability of profile \eqref{choice0}, as this will set the basis for future computations. 
Applying the Cauchy-Schwarz inequality we have 
\begin{eqnarray*}
  \Q\{\theta\}&=&\lla\int_0^1|\nabla\theta|^2\, dz-\frac{2}{\delta}\int_0^{\delta}\uz\theta\, dz\rra\\
  &\geq&\lla\int_0^1|\nabla\theta|^2\, dz\rra-\frac{2}{\delta}\lla\int_0^{\delta}|\uz|^2\, dz\rra^{\frac 12}\lla\int_0^{\delta}|\theta|^2\, dz\rra^{\frac 12}\,.
\end{eqnarray*}
Thanks to the boundary conditions for $\uz$ and $\theta$, we can use the Poincar\'e estimates 
\begin{equation}\label{pc-u}
  \lla\int_0^{\delta}|\uz|^2\, dz\rra\leq\delta \lla\int_0^{\delta}|\partial_z\uz|^2\, dz\rra\,,
\end{equation}
\begin{equation}\label{pc-theta}
  \lla\int_0^{\delta}|\theta|^2\, dz\rra\leq\delta \lla\int_0^{\delta}|\partial_z\theta|^2\, dz\rra
\end{equation}
together with \eqref{Nu-def2} and \eqref{diss-bound} to get
$$\Nu(1-\delta\Ra^{\frac 12})\geq 0\,. $$
Optimizing on the boundary layers thickness, we select
$$\delta\sim \Ra^{-\frac 12}\,,$$
yielding
$$\Nu\leq \int_0^1|\tau'|^2\, dz\leq \frac{1}{\delta}\sim\Ra^{\frac 12}\,.$$
Therefore we proved 
\begin{thm}\label{th1}
For solutions of \eqref{RBC}--\eqref{Tbc} with no-slip boundary conditions \eqref{no-slip} the following upper bound holds:
$$\Nu\lesssim \Ra^{\frac 12} \qquad \mbox{ for }\quad \Pra=\infty\,.$$
\end{thm}
We remark that, in order to obtain this bound we could equivalently use the localization principle in \eqref{av-Nusselt}, thus avoiding going through the background field method. In fact by the Cauchy-Schwarz inequality and estimates \eqref{pc-u} and \eqref{pc-theta}, we obtain
\begin{eqnarray*}
  \Nu&\leq&\frac{1}{\delta}\lla\int_0^{\delta}  u^zT\, dz\rra+\frac{1}{\delta}\\
  &\leq&\delta \lla\int_0^{\delta} |\partial_z u^z|^2\, dz\rra^{\frac 12}\lla\int_0^{\delta} |\partial_z T|^2\, dz\rra^{\frac 12}+\frac{1}{\delta}\\
  &\leq& \delta(\Nu-1)^{\frac 12}\Ra^{\frac 12}\Nu^{\frac 12}+\frac{1}{\delta}\\
  &\lesssim& \delta\Nu\Ra^{\frac 12}+\frac{1}{\delta}\,. 
\end{eqnarray*}
Balancing, the optimal $\delta$ satisfies
$$\delta^2\sim\frac{1}{\Nu\Ra^{\frac 12}}\,,$$
leading to
$$\Nu\lesssim \Ra^{\frac 12}\,.$$
Therefore the localization method is equivalent to the background field method with the choice \eqref{choice0}.
More discussions and interesting observations about the relation between the background field method (and auxiliary functional method) and the localization method may be found in \cite{C17}.
\paragraph{Blooming of the background field method:}
In \cite{DC01}, Doering and Constantin choose (a smooth approximation of) the profile 
\begin{equation}\label{choice2}
\tau(z)=
\begin{cases}
  1-\frac{z}{2\delta} & \quad 0\leq z\leq \delta\\
  \frac 12 & \quad \delta\leq z\leq 1-\delta\\
  \frac{1-z}{2\delta} & \quad 1-\delta\leq z\leq 1\,,
\end{cases}
\end{equation}
and improve Theorem \ref{th1} by employing finer estimates for the vertical velocity $\uz$. Here we describe their argument. 
\begin{figure}[h!]
\centering
\includegraphics[scale=0.5]{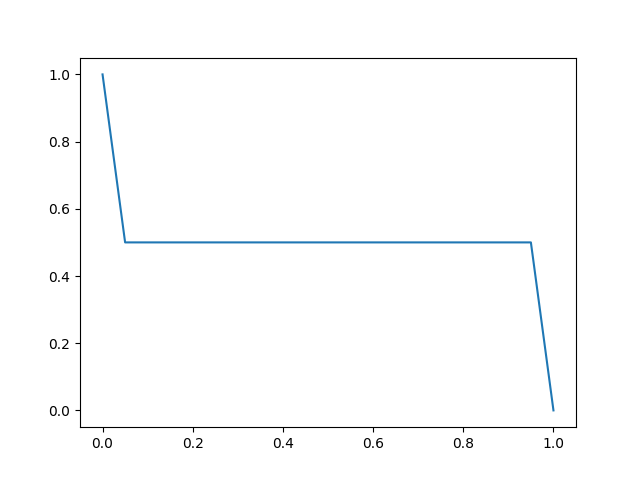}
\caption{Plot of the background profile $\tau=\tau(z)$ used in \cite{DC01}.}
\end{figure}
Thanks to periodicity in the horizontal directions, passing to Fourier-series offers a big advantage. In fact, the unknown functions can be written as Fourier series with $z-$ dependent coefficients \footnote{Here we assume, for simplicity $L=L_x=L_y$.}
$$ \vv(\xp,z)=\sum_{\mathbf{j}\in \mathbb{Z}^2}\hat{\vv}(z)e^{\frac{2 \pi i}{L}(\xp\cdot \mathbf{j})}\,.$$
Then the equation for \eqref{4order-elliptic} decomposes in 
\begin{equation}\label{4order-FT}
  \left(-\frac{d^2}{dz^2}+k^2\right)^2\hatuz=\Ra k^2\hatt\,,
\end{equation}
where $k=\frac{2\pi}{L}|\mathbf{j}|$ and the
boundary conditions for $\uz$ are
\begin{equation}\label{bc-fourth}
  \hatuz(0)=0=\hatuz(1) \quad \mbox{ and } \quad \partial_z\hatuz(0)=0=\partial_z\hatuz(1)\,.
\end{equation}
Decomposing $\Q$ mode by mode in the translation invariant horizontal directions, it is clear that $\Q$ will be non-negative, if for each horizontal 
wavenumber $k$ 
\begin{equation}\label{quad-form-in-k}
\hat\Q\{\hatt\}=\int_0^1 \left( |\partial_z\hatt|^2+k^2|\hatt|^2+\tau'(\hatuz^{\ast}\hatt +\hatuz \hatt^{\ast})\right)\, dz\,,
\end{equation}
is non-negative. Here the symbol $^{*}$ denotes the complex conjugate.
Because of the choice of the background profile in \eqref{choice2} we have 
\begin{equation*}
\int_0^1 \tau'(\hatuz^{\ast}\hatt +\hatuz \hatt^{\ast})\, dz\leq \frac{1}{\delta}\left(\int_0^{\delta}|\hatuz(z)||\hatt(z)|\, dz+ \int_{1-\delta}^{1}|\hatuz(z)||\hatt(z)|\, dz\right)\,.
\end{equation*}
The crucial element leading to the new bound is contained in the way the authors estimate the mixed term. To illustrate the argument, let us focus on the lower boundary layer. By Poincar\'e's inequality applied to $\uz$, $\theta$ (see \eqref{pc-u} and \eqref{pc-theta}) and $\partial_z \uz$ (thanks to the fact that $\partial_z\uz=0$ at $z=\{0,1\}$)  we have 
\begin{eqnarray}
\frac{1}{\delta}\int_0^{\delta}|\hatuz(z)||\hatt(z)|\, dz&\leq& 
\frac{1}{\delta}\left(\int_0^{\delta}|\hatuz(z)|^2\, dz\right)^{\frac 12}\left(\int_0^{\delta}|\hatt(z)|^2\, dz\right)^{\frac 12}\notag\\
&\leq&\frac{1}{\delta}\delta^2\left(\int_0^{\delta}|\partial_z^2\hatuz(z)|^2\, dz\right)^{\frac 12}\delta\left(\int_0^{\delta}|\partial_z\hatt(z)|^2\, dz\right)^{\frac 12}\notag\\
&\leq&\delta^{\frac 52}\sup_{z}|\partial_z^2\hatuz(z)|\left(\int_0^1|\partial_z\hatt(z)|^2\, dz\right)^{\frac 12}\,.\label{est-mt}
\end{eqnarray}
Again, due to the boundary conditions $\uz=0$ and $\partial_z\uz=0$ at $z=\{0,1\}$, the $L^{\infty}$ norm of $\uz$ can be bounded by using the following interpolation estimate
\begin{lemma}
Let $v(z):[0,1]\rightarrow \R$ be a smooth function satisfying $v=0$ and $\partial_z v=0$ at $z=\{0,1\}$, then 
\begin{equation}\label{interpol}
\|\partial_z^2v\|_{L^{\infty}(0,1)}^2\leq2\|\partial_z^4v\|_{L^2(0,1)}\|\partial_z^2v\|_{L^2(0,1)}\,.
\end{equation} 
\end{lemma}

Applying this lemma to $v=\uz$ and using Young's inequality we obtain
\begin{equation*}
\|\partial_z^2\hatuz\|_{L^{\infty}(0,1)}^2\leq k^2\|\partial_z^2\uz\|_{L^2(0,1)}^2+\frac{1}{k^2}\|\partial_z^4\uz\|_{L^2(0,1)}^2\,.
\end{equation*}
In order to conclude the argument, we now reconnect to the equation.
In fact, thanks to the instantaneous slaving in \eqref{4order-FT} one can easily deduce 
$$k^2\|\partial_z^2\hatuz\|_{L^2(0,1)}^2+\frac{1}{k^2}\|\partial_z^4\hatuz\|_{L^2(0,1)}^2\leq \frac 1C \Ra^2k^2\|\hatt\|_{L^2(0,1)}^2\,,$$
for some numerical constant $C>0$.
In fact, squaring \eqref{4order-FT} and integrating by parts we have 
\begin{eqnarray*}
  \Ra^2 k^4\|\hatt\|_{L^2(0,1)}^2&=&\|\partial_z^4\hatuz+k^4\hatuz-2k^2\partial_z^2\hatuz\|_{L^2(0,1)}^2\\
  &\geq&\|\partial_z^4\hatuz\|_{L^2(0,1)}^2+k^8\|\hatuz\|_{L^2(0,1)}^2+ 4k^4\|\partial_z^2\hatuz\|_{L^2(0,1)}^2\\
  &&+ 2k^4\int_0^1 \partial_z^4\hatuz \,\hatuz\, dz-4k^6\int_0^1\hatuz \,\partial_z^2\hatuz\, dz-
    4 k^2\int_0^1|\partial_z^4\hatuz\,\partial_z^2\hatuz|\, dz\\
  &=&\|\partial_z^4\hatuz\|_{L^2(0,1)}^2+k^8\|\hatuz\|_{L^2(0,1)}^2+ 6k^4\|\partial_z^2\hatuz\|_{L^2(0,1)}^2
      + 4k^6\|\partial_z\hatuz\|_{L^2(0,1)}^2-4 k^2\int_0^1|\partial_z^4\hatuz|\,|\partial_z^2\hatuz|\, dz\\
  &\geq&\|\partial_z^4\hatuz\|_{L^2(0,1)}^2+ 6k^4\|\partial_z^2\hatuz\|_{L^2(0,1)}^2
        -a\|\partial_z^4\hatuz\|_{L^2(0,1)}^2-\frac{4 k^4}{a}\|\partial_z^2\hatuz\|_{L^2(0,1)}^2\,.
\end{eqnarray*}
Choosing $a=\frac{7}{10}$, we obtain the desired bound.
Finally going back to \eqref{est-mt} we have
\begin{eqnarray*}
  \frac{1}{\delta}\int_0^{\delta}|\hatuz(z)||\hatt(z)|\, dz&\leq& 
  \delta^{\frac 52}\sup_{z}|\partial_z^2\hatuz(z)|\left(\int_0^1|\partial_z\hatt(z)|^2\, dz\right)^{\frac 12}\,\\
  &\leq&  \frac{\delta^{\frac 52}}{\sqrt{C}} \Ra k\|\hatt\|_{L^2(0,1)}\|\partial_z \hatt\|_{L^2(0,1)}\\
  &\leq &\frac{\delta^{\frac 52}}{2\sqrt{C}}\Ra (k^2\|\hatt\|_{L^2(0,1)}^2+\|\partial_z \hatt\|_{L^2(0,1)}^2)\,.
\end{eqnarray*}
Then, from \eqref{quad-form-in-k} 
\begin{equation*}
  \hat\Q\{\hatt\}\geq 
  \left(1-\frac{\delta^{\frac 52}}{2\sqrt{C}}\Ra \right)(\|\partial_z\hatt\|^2_{L^2}+k^2\|\hatt\|_{L^2(0,1)})\,.
\end{equation*}
The choice $\delta\sim \Ra^{-\frac 25}$ in \eqref{Denergy} yields
\begin{thm}\label{th2}
For solutions of \eqref{RBC}--\eqref{Tbc} with no-slip boundary conditions \eqref{no-slip} the following upper bound holds:
$$\Nu\lesssim \Ra^{\frac 25} \qquad \mbox{ for }\quad \Pra=\infty\,.$$
\end{thm} 
From the previous argument we see that the choice of bounding the mixed term with the $L^{\infty}-$norm of the second derivative of $\uz$ ``maximizes'' the power of $\delta$. In fact this last method gives us a $\delta^{\frac 52}$ prefactor against the $\delta^2$ prefactor of the previous section. Moreover notice that in this proof, the instantaneous slaving of $\uz$ to $\theta$ given by \eqref{4order-elliptic} was crucial.
It is now natural to ask whether the profile chosen in \eqref{choice2} is optimal. Before jumping to the next two seminal results, let us comment on the background profile chosen in \eqref{choice2}: The temperature field is supposed to be laminar in very thin boundary layers and this justifies the linear profile, decreasing from $z=0$ to $z=\delta$ and from $z=1-\delta$ to $z=1$. In the bulk instead the temperature is not expected to vary wildly.
\newline
Although the profile in \eqref{choice2} turned out to be \textit{marginally} stable for some choice of very small $\delta$, it does not
reflect the idea that the fluid layers must be ``stably stratified''. This means that the lighter, warmer parcels of fluids are expected to be on the top of the cold, denser parcels. This observation leads us to the next result.

\paragraph{Almost crowning of the background field method:}
In \cite{DOR06}, Doering, Otto and Westdickenberg (ne\'e Reznikoff) show stability of the logarithmic background profile
\begin{equation}\label{choice3}
\tau(z)=
\begin{cases}
1-\frac{z}{\delta} & \quad 0\leq z\leq \delta\\
\frac 12+\lambda(\delta)\ln\left(\frac{z}{1-z}\right) & \quad \delta\leq z\leq 1-\delta\\
\frac{(1-z)}{\delta} & \quad 1-\delta\leq z\leq 1\,,
\end{cases}
\end{equation}
with 
\begin{equation}\label{lambda}
\lambda(\delta)=\frac{1}{2\ln((1-\delta)/\delta)}.
\end{equation}
\begin{figure}[h!]
\centering
\includegraphics[scale=0.5]{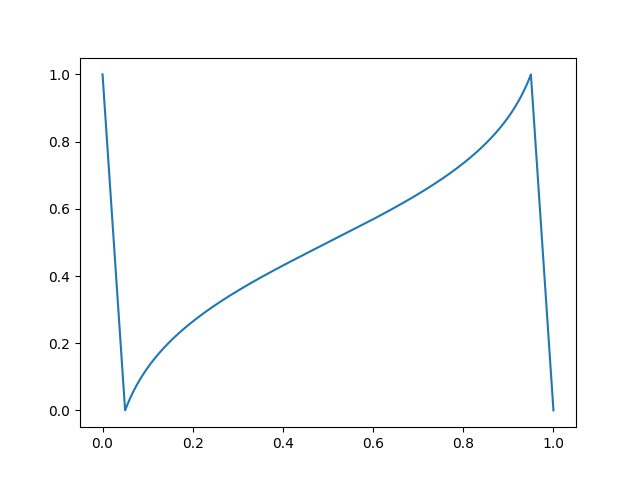}
\caption{Plot of the background profile $\tau=\tau(z)$ chosen in \cite{DOR06}.}
\end{figure}
In this case, the derivative of the background profile produces weighted terms which are subtler to control, as we are going to describe now.
With the choice \eqref{choice3}, the quadratic form becomes
\begin{equation}\label{Q-dec}
\Q\{\theta\}=\int_0^1  \la|\nabla \theta|^2\ra\, dz
+2\lambda(\delta)\int_{\delta}^{1-\delta}\left(\frac 1z+\frac{1}{1-z}\right)\; \la\theta\uz\ra\, dz
-\frac{1}{\delta}\int_{0}^{\delta} \la\theta \uz\ra \, dz\,
-\frac{1}{\delta}\int_{1-\delta}^{1}  \la \uz\theta\ra \, dz\,.
\end{equation}
Adding and subtracting $(\frac{\lambda}{z}+\frac{\lambda}{1-z})\la\uz\theta\ra$ in the boundary layers we rewrite
\begin{equation*}
\Q\{\theta\}=\Q\{\theta\}_{\rm{T}}+\Q\{\theta\}_{\rm{B}}\,,
\end{equation*}
where 
\begin{eqnarray}\label{Q_B}
\Q_{\rm{B}}\{\theta\}=\int_{0}^{\frac 12}  \la|\nabla \theta|^2\ra\, dz
+2\lambda(\delta)\int_{0}^{1}\frac{\la\uz\theta\ra}{z}\, dz
-\int_{0}^{\delta}\left(\frac{1}{\delta}+\frac{\lambda(\delta)}{z}+\frac{\lambda(\delta)}{1-z}\right) \la\uz\theta\ra\, dz\,.
\end{eqnarray}
and
\begin{eqnarray*}
\Q_{\rm{T}}\{\theta\}=\int_{1/2}^1  \la|\nabla \theta|^2\ra\, dz
+2\lambda(\delta)\int_{0}^{1}\frac{\la\uz\theta\ra}{1-z}\, dz
-\int_{1-\delta}^{1}\left(\frac{1}{\delta}+\frac{\lambda(\delta)}{z}+\frac{\lambda(\delta)}{1-z}\right) \la\uz\theta\ra\, dz\,.
\end{eqnarray*}
Smuggling in the weight $z^2$ and applying H\"older and Cauchy-Schwarz inequalities, the last term in $\Q_{\rm{B}}$ can be estimated as follows:
\begin{eqnarray*}
  &&\int_{0}^{\delta}\left(\frac{1}{\delta}+\frac{\lambda(\delta)}{z}+\frac{\lambda(\delta)}{1-z}\right) \la\uz\theta\ra\, dz\\
  &&=\int_{0}^{\delta}\left(\frac{1}{\delta}+\frac{\lambda(\delta)}{z}+\frac{\lambda(\delta)}{1-z}\right) \;z^2 \;\frac{\la|\theta|^2\ra^{\frac 12}}{z^{\frac 12}}\;\frac{\la|\uz|^2\ra^{\frac 12}}{z^{\frac 32}}\, dz\\
&&\leq\left(\sup_{0<z<\frac 12}\frac{\la|\theta|^2\ra^{\frac12}}{z^{\frac 12}}\right)\left(\int_0^1\frac{\la|\uz|^2\ra}{z^3}\, dz\right)^{\frac 12}\left(\int_0^{\delta}z^4\left[ \frac{1}{\delta}+\frac{\lambda}{z}+\frac{\lambda}{1-z}\right]^2\, dz\right)^{\frac 12}\,.
\end{eqnarray*}
By the fundamental theorem of calculus  
$$\sup_{0\leq z\leq \delta}\frac{\la|\theta|^2\ra}{z}\leq \int_0^1\la|\partial_z\theta|^2\ra\, dz\leq \int_0^1\la|\nabla\theta|^2\ra\, dz\,, $$
and, by a direct computation
$$\int_0^{\delta}z^4\left(\frac{1}{\delta}+\frac{\lambda}{z}+\frac{\lambda}{1-z}\right)^2\, dz\sim \delta^3\,.$$
The crucial estimate in this paper is contained in the following
\begin{lemma}
For $\theta\in L^2(\Omega)$ and $w$ satisfying \eqref{4order-FT} and \eqref{bc-fourth}, the bound
\begin{equation}\label{weighted-es}
\int_0^1\frac{ \la w\theta \ra}{z}\, dz\gtrsim \frac{1}{\Ra}\int_0^1\frac{\la|w|^2\ra}{z^3}\, dz\,.
\end{equation}
holds.
\end{lemma}
The combination of this lemma applied to $w=\uz$ with the estimates above yields
$$\Q_{\rm{B}}\{\theta\}\gtrsim\left[\frac{2\lambda}{\Ra}-\delta^3\right]\int_0^1\frac{\la|\uz|^2\ra}{z^3}\, dz\,.$$
Arguing in the same way for $\Q_{\rm{T}}$ and imposing $\delta^3\ln((1-\delta)/\delta)=\frac{1}{2\Ra}$, one finds that $\delta\sim \frac{1}{\Ra\ln\Ra}$ is optimal, and, since $\Nu\leq \int_0^1(\tau')^2\, dz\sim \frac{1}{\delta}$, this proves the following 
\begin{thm}
For solutions of \eqref{RBC}--\eqref{Tbc} with no-slip boundary conditions \eqref{no-slip} the following upper bound holds:
\begin{equation}\label{triangle}
\Nu\lesssim\Ra^{\frac 13}(\ln\Ra)^{\frac 13} \qquad \mbox{ for } \quad \Pra=\infty\,.
\end{equation}
\end{thm}
This result improves Theorem \ref{th2} and was the first one capturing (with an upper bound) Malkus' scaling $\Ra^{\frac 13}$ \textit{up to a logarithmic correction} only using the background field method. The choice of the logarithmic background
reveled itself to be ``optimal'' in a sense that we will specify at the end of this section. Roughly speaking, the logarithmic profile respects the stable stratification since it (monotonically) grows in the bulk. And the slow growth in the bulk maintains stability. 

\paragraph{Improving the (logarithmic) failure:}

The logarithmic correction in \eqref{triangle} was later improved by Otto and Seis in \cite{OS11} using the following idea: going back to \eqref{Q-dec}
and focusing on \eqref{Q_B}, the authors proved that, for $(\lambda\Ra)^{-\frac 13}\leq1$ the upper bound
\begin{equation}\label{crucial-es}
\sup_{0\leq z\leq (\lambda\Ra)^{-\frac 13}}\langle|\partial_z^2 \uz|^2\rangle\leq \Ra^{\frac 53}\lambda^{-\frac 13}\left[\int_0^1\lla |\nabla \theta|^2\rra \, dz+\lambda \int_0^1\frac{\lla\theta \uz\rra}{z}\right]
\end{equation}
holds. Estimate \eqref{crucial-es} is crucial for reducing the power of $\lambda$ in the bound: in fact, applying it in the estimate
\begin{eqnarray*}
\left|\int_0^\delta\left(\frac{1}{\delta}+\frac{\lambda}{z}+\frac{\lambda}{1-z}\right)\la \theta\uz\ra\, dz\right|
&\lesssim& \sup_{0\leq z\leq\delta}\frac{\langle |\theta|^2\rangle^{\frac 12}}{z^{\frac12}}\sup_{0\leq z\leq \delta}\frac{\langle |\uz|^2\rangle^{\frac 12}}{z^2}\int_0^{\delta}\left(\frac{1}{\delta}+\frac{\lambda}{z}+\frac{\lambda}{1-z} \right)z^{\frac 52}\, dz\,\\
&\lesssim&\delta^{\frac 52}\left(\int_0^{\delta}\langle|\partial_z\theta|^2\rangle\, dz\right)^{\frac 12} \;\sup_{0\leq z\leq \delta}\langle|\partial_z^2\uz|^2\rangle^{\frac 12}\\
&\lesssim& \delta^{\frac 52}\left(\int_0^{\delta}\langle|\partial_z\theta|^2\rangle\, dz\right)^{\frac 12}\; \Ra^{\frac 56}\lambda^{-\frac 16}\left[\int_0^1\lla |\nabla \theta|^2\rra \, dz+\lambda \int_0^1\frac{\lla\theta \uz\rra}{z}\right]^{\frac 12}\,,
\end{eqnarray*}
where we used the Poincar\'e-type inequality
$$\sup_{0\leq z\leq \delta}\frac{\lla|\uz|^2\rra}{z^4}\leq \sup_{0\leq z\leq \delta}\langle|\partial_z^2\uz|^2\rangle\,,$$
the authors then conclude
\begin{eqnarray*}
\Q_{\rm{B}}\{\theta\}&\geq& \frac 12\int_0^1\lla |\nabla \theta|^2\rra \, dz+2\lambda \int_0^1\frac{\lla\uz\theta\rra}{z}\, dz-\left|\int_{0}^{\delta}\left(\frac{1}{\delta}+\frac{\lambda}{z}+\frac{\lambda}{1-z}\right)\lla \theta\uz\rra\, dz\right|\\
&\geq&\frac 12\int_0^1\lla |\nabla \theta|^2\rra \, dz+2\lambda \int_0^1\frac{\lla \uz\theta\rra}{z}\, dz\\
&&-\delta^{\frac 52}\left(\int_0^{\delta}\lla|\nabla\theta|^2\rra\, dz\right)^{\frac 12}\; \Ra^{\frac 56}\lambda^{-\frac 16}2\left[\frac 14\int_0^1\lla |\nabla \theta|^2\rra \, dz+\frac{\lambda}{4} \int_0^1\frac{\lla\theta \uz\rra}{z}\, dz\right]^{\frac 12}\\
&\geq&\frac 12\int_0^1\lla |\nabla \theta|^2\rra \, dz+2\lambda \int_0^1\frac{\lla\theta \uz\rra}{z}\, dz\\
&&-2\delta^{\frac 52}\Ra^{\frac 56}\lambda^{-\frac 16}
\left\{
      \varepsilon\int_0^{1}\lla|\nabla\theta|^2\rra\,dz+\frac{1}{4\varepsilon}
      \left[\frac 14\int_0^1\lla |\nabla \theta|^2\rra \, dz+\frac{\lambda}{4} \int_0^1\frac{\lla\theta \uz\rra}{z}\, dz
      \right]
\right\}\,.
\end{eqnarray*}
Choosing $\varepsilon=\frac 14$ and using $\int_0^1\frac{\lla\theta \uz\rra}{z}\, dz\geq 0$ by \eqref{weighted-es}, we  have 
\begin{eqnarray*}
  \Q_{\rm{B}}\{\theta\}
  &\geq&\frac 12\int_0^1\lla |\nabla \theta|^2\rra \, dz+2\lambda \int_0^1\frac{\lla\theta \uz\rra}{z}\, dz\\
  &&-\delta^{\frac 52}\Ra^{\frac 56}\lambda^{-\frac 16}
  \left\{\frac 12\int_0^1\lla |\nabla \theta|^2\rra \, dz+2\lambda \int_0^1\frac{\lla\theta \uz\rra}{z}\, dz
  \right\}\\
  &=&
  (1-\delta^{\frac 52}\Ra^{\frac 56}\lambda^{-\frac 16})
  \left\{\frac 12\int_0^1\lla |\nabla \theta|^2\rra \, dz+2\lambda \int_0^1\frac{\lla\theta \uz\rra}{z}\, dz
  \right\}\,.
\end{eqnarray*}
The optimal $\delta$ is
$$\delta\sim \Ra^{-\frac 13}(\ln\Ra)^{-\frac{1}{15}}\,,$$
and, as a result, we showed
\begin{thm}
For solutions of \eqref{RBC}--\eqref{Tbc} with no-slip boundary conditions \eqref{no-slip} the following upper bound holds:
$$\Nu\lesssim \Ra^{\frac 13}(\ln\Ra)^{\frac{1}{15}}\qquad \mbox{ for } \quad \Pra=\infty\,.$$
\end{thm}
The new estimate for the velocity in \eqref{crucial-es} improves the logarithmic correction substantially, but the bound still fails to reproduce the classical scaling $\Ra^{\frac 13}$. It is now natural to ask
\begin{itemize}
\item can we find a ``better'' profile which captures the Malkus' scaling without corrections? or
\item is it possible to refine the regularity estimates to further reduce the logarithmic correction?
\end{itemize}
We will (partially) answer these questions in the next sections.

\paragraph{Limitations of the background field method:}
As we have seen in the previous section, the background field method transforms the problem of finding bounds on the Nusselt number into a variational problem. If we would find the minimizer of this problem, we would get the optimal upper bound that this method is able to produce. 
In particular, the shape of the optimal $\tau$ would give us precise information about the (statistical) behavior of the temperature. 
\newline
Let us rescale the infinite Prandtl number system according to the rescaling \eqref{rescaling1} suggested by Malkus’ marginal stability argument, and set $H=\Ra^{\frac 13}$ to obtain 
\begin{equation*}
\begin{array}{rclcl}
  \partial_{ t}  T+ \uu\cdot \nabla  T &=& \Delta T &\qquad 0<z<H \\
    -\Delta \uu+\nabla p&=& T e_{ z}  &\qquad 0<z<H \\
    \nabla\cdot \uu&=& 0  &\qquad 0<z<H \\
    \uu&=&0 &\qquad z\in \{0,H\}\\
    T&=&1 &\qquad z=0\\
    T&=&0 &\qquad z=H\,.\\
\end{array}
\end{equation*}
Let us define the Nusselt number associated to the background flow method
\begin{equation}\label{min-max}
\Nubf= \inf_{\substack{\tau:[0,H]\rightarrow \R:\\\tau(0)=1\;,\tau(H)=0} }\left\{\int_0^H|\tau'(z)|^2\, dz-\Q\{\theta\}  \right\}
\end{equation}
with $\Q$ being
\begin{equation}\label{stab-con}
\Q\{\theta\}=\lla\int_0^H|\nabla \theta|^2\, dz+2\int_0^H\tau'\uz\theta\, dz \rra\,,
\end{equation}
where $\theta$ is instantaneously ``slaved'' to $\uz$ through 
 \begin{equation*}
 -\Delta^2 \uz= \DeltaH\theta \quad 0<z<H, \qquad \uz=\partial_z\uz=0\quad z\in\{0,H\}\,.
 \end{equation*}
Clearly, 
$$\Nu\leq\Nubf\,,$$
and the challenge is to understand how far the Nusselt number $\Nu$ is from the number $\Nubf$ produced by the background field method.
\newline 
In \cite{NO17} the authors prove 
\begin{thm}
For solutions of \eqref{RBC}--\eqref{Tbc} with no-slip boundary conditions \eqref{no-slip}, $\Pra=\infty$ and $\Ra\gg 1$ the following upper bound holds: 
\begin{equation}\label{NO-result}
\Nubf\gtrsim (\ln\Ra)^{\frac{1}{15}}\Ra^{\frac 13}\,.
\end{equation}
\end{thm}
This ansatz-free lower bound is proven by extracting local information on $\tau$ from the non-local stability condition \eqref{stab-con}, which written in Fourier-series reads:
\begin{eqnarray*}
2\int_0^H \tau'\hat{\uz}\left(-\frac{d^2}{dz^2}+|{\bf{k}}|^2\right)^2\hat{\uz}^{\ast}\, dz
+\int_0^H|{\bf{k}}|^{-2}\left|\frac{d}{dz}\left(-\frac{d^2}{dz^2}+|{\bf{k}}|^2\right)^2\hat{\uz}\right|^2\, dz
+\int_0^H\left|\left(-\frac{d^2}{dz^2}+|{\bf{k}}|^2\right)^2\hat{\uz}\right|^2\, dz\geq 0\,,
\end{eqnarray*}
with ${\bf{k}}\in \frac{2\pi}{L}\mathbb{Z}^{d-1}\setminus\{0\}$ and 
\begin{equation}\label{bc-f}
\hat{\uz}=\frac{d}{dz}\hat{\uz}=\left(-\frac{d^2}{dz^2}+|{\bf{k}}|^2\right)^2\hat{\uz}=0\quad \mbox{ for } z\in\{0,H\}\,.
\end{equation}
Notice that in the quadratic form only the modulus of $k$ appears, so that the stability condition is independent of the dimension.
The authors initially show the result in a simplified setting.
First, considering profiles $\tau$ that are stable even
under perturbation that have horizontal wave-length much larger than $H$, letting the side-length $L\rightarrow \infty$, implying $k\in \R$.  
Second, reducing the stability condition to the second term in \eqref{stab-con}, which is the leading order term in the stability condition. Under these assumptions the authors characterize all stable profiles:
\begin{prop}
 If $\frac{d\tau}{dz}$ satisfies 
\begin{equation}\label{RSC}
\int_0^H \tau'\hat{\uz}\left(-\frac{d^2}{dz^2}+|k|^2\right)^2\hat{\uz}^{\ast}\, dz\geq 0
\end{equation}
for all $k\in \R$ and all $\hat{\uz}$ satisfying \eqref{bc-f}, then 
\begin{eqnarray*}
\tau'\geq 0
\end{eqnarray*}
and 
$$\int_{\frac 1e}^{1}\tau'\, dz\gtrsim \frac{1}{\ln H}\int_{1}^{H}\tau'\, dz\,.$$
\end{prop}
This means that, if $\tau$ satisfies the reduced stability condition \eqref{RSC}, then it must be increasing and ``logarithmically'' growing in the bulk.
We remark that, in \cite{Nobili-thesis} it is proved that the result above still holds true if the lateral size $L$ is of order $H$. Extending this result to the full stability condition requires some careful analysis: by
subtle averaging of the stability condition, the authors in \cite{NO17} construct a non-negative convolution kernel with
the help of which they can express the positivity on average approximately in the bulk.
Also the logarithmic growth can be recovered at least approximately in the bulk. To finally achieve the lower bound, further estimates connecting the bulk with the boundary layers are provided.
This result proves the optimality of the logarithmic profile.
\newline
Interestingly, by a mixture of analytical and numerical analysis, Ierely, Kerswell and Plasting in 2006 \cite{IKP06} anticipated (although slightly underestimating) the result in \cite{NO17}, showing $\Nu\sim \Ra^{0.33173}(\ln\Ra)^{0.0325}$.

\subsubsection{The Constantin\&Doering '99 argument and the best upper bound at infinite Prandtl number}\label{CD99}
In 1999 Constantin \& Doering \cite{CD99} caught the Malkus scaling $\Ra^{\frac 13}$ (up to a logarithm) without optimizing on the background profiles, but extracting finer information from the equation for the velocity. This paper marks the beginning of a new and powerful way of thinking about this problem. 
The authors choose $\tau$ to be (a smooth approximation of) the profile
\begin{equation}\label{CD-profile}
\tau=
\begin{cases}
1-\frac{z}{\delta} & 0\leq z\leq \delta\\
0 & z\geq \delta\,,
 \end{cases}
 \end{equation}
 in 
 $$\Nu=\lla\int_0^1|\tau'|^2\,dz \rra-\lla\int_0^1|\nabla\theta|^2\, dz
 +2\int_0^1\tau'\uz \theta\, dz\rra\,.$$
 An application of the maximum principle $\|\theta\|_{L^{\infty}(\Omega)}\leq 1$ to estimate the mixed term
 \begin{eqnarray*}
\left|2\fint_{\Omega}\tau'\uz \theta\, dz\,dy\, dx\right|
 &\lesssim&\|\partial_z^2 \uz\|_{L^{\infty}((0,1); L^1([0,L_x]\times[0,L_y]))}\,\int_0^{\delta}|\tau'|\, z^2\, dz\,
 \end{eqnarray*}
 yields
 \begin{equation}\label{avbound}
 \Nu\lesssim \delta^2 \lla\sup_{0\leq z\leq 1}|\partial_z^2 \uz|\, dz\rra+\frac{1}{\delta}\,, 
 \end{equation}
where the bound $|\tau'(z)|\lesssim \frac{1}{\delta}$ was used.

As we have seen in Section \ref{total}, the interpolation estimate \eqref{interpol} was the crucial ingredient for proving the bound $\Nu\lesssim \Ra^{\frac 25}$, but turned out to be sub-optimal. In this paper the authors proved the much finer estimate
\begin{prop}
Suppose $w$ solves the problem 
\begin{equation*}
\begin{array}{rlll}
-\Delta^2w&=&\DeltaH f \quad  & \mbox{ in }[0,L]^2\times[0,1]\\
w=\partial_z w&=&0 \quad  & \mbox{ at } z=\{0,1\}
\end{array}
\end{equation*}
with horizontally periodic boundary conditions. 
Then, for any $\alpha\in (0, 1)$ there exists a positive constant $C_{\alpha}$
such that the bound
\begin{equation}\label{max-reg1}
\|\partial_z^2w\|_{L^{\infty}(\Omega)}\leq C_{\alpha}\|f\|_{L^{\infty}(\Omega)}(1+\log_{+} \|f\|_{C^{0,\alpha}(\Omega)}))^2\,,
\end{equation}
holds, where 
$$\|f\|_{C^{0,\alpha}(\Omega)}=\sup_{\xx\in \Omega}|f(x,t)|+\sup_{x\neq y}\frac{|f(x,t)-f(y,t)|}{|x-y|^{\alpha}}\,.$$
\end{prop} 
This \textit{maximal regularity estimate} was obtained
by decomposing the operator $B=\partial_z^2(\Delta^2)^{-1}\DeltaH$ into the sum of non translationally invariant operators, whose kernel is not explicit. 
Inserting the long-time average of \eqref{max-reg1} into \eqref{avbound} and using 
\begin{equation}\label{est-Dtheta}
\lla\int_0^1|\Delta\theta|^2\, dz\rra\leq C\Ra^2\left\{1+\int_0^1[|\tau''(z)|^2+ \;z\;|\tau'(z)|^2]\, dz\right\}\,,
\end{equation}
to estimate the $C^{0,\alpha}$ norm in the logarithmic correction further
\footnote{This estimate is simply obtained by testing \eqref{flactuation-equation} with $\Delta \theta$, integrating by parts, and using the Cauchy-Schwartz and Youngs inequalities together with the interpolation inequality
$\|\nabla \theta\|_{L^4(\Omega)}^2\leq C\|\theta\|_{L^{\infty}(\Omega)}\|\Delta \theta\|_{L^2(\Omega)}\,$
and 
$\|\nabla\uu\|_{L^2(\Omega)}^2\leq C\Ra^2$, where the last inequality is derived by testing the Navier-Stokes equations, integrating by parts and using the maximum principle for the temperature.},
the authors obtain
$$
\Nu\lesssim \delta^2\Ra(1+\log(\Ra))^2+\frac{1}{\delta}
$$
 and the choice $\delta\sim \Ra^{-\frac 13}(1+\log_+\Ra)^{-\frac 23}$ yields
$$\Nu\leq \Ra^{\frac 13}(1+\log_+\Ra)^{\frac 23}\,.  $$
We notice that the logarithmic correction in this bound is bigger than the one produced in \cite{OS11} using the logarithmic profile in the background field method. Nevertheless, we are now going to discuss that, by a refinement of this argument, Otto and Seis in \cite{OS11} proved the best upper bound (up to today). 
We first observe that, by the maximum principle and Poincar\'e's inequality (applied twice in $z$) the bound
\begin{eqnarray*}
\Nu&\leq& \frac{1}{\delta}\lla \int_0^1\theta\uz\, dz\rra+\frac{1}{\delta}\\
&\leq&\delta^2\sup_z\langle|\partial_z^2\uz|\rangle^{\frac 12}+\frac{1}{\delta}
\end{eqnarray*}
holds and is the same as \eqref{avbound} (which was obtained by \eqref{av-Nusselt}).
 The key of the result in \cite{OS11} is the estimate on the hessian of $\uz$
\begin{equation}\label{toinsert}
\sup_{z\in (0,1)}\lla|\nabla^2\uz|^2\rra\lesssim \Ra^{2}\left( \Ra^{-\frac 13}\Nu+\ln^2(\Ra^{-\frac 13}\Nu\ln^{\frac 12}(\Ra^{\frac 13}))\right)
\end{equation}
obtained by a clever combination of the \textit{maximal regularity estimate}
\begin{equation}\label{max-reg2}
\lla\sup_{z\in (0,1)}|\nabla^2\uz|^2\rra^{\frac 12}
\lesssim \Ra\ln\left(\frac{\lla\int_{0}^{1}|\nabla \theta|^2\, dz\rra}{\lla\sup_{\xx}\theta^2\rra}+e\right)\la\sup_{\xx}\theta^2\ra^{\frac 12}
\end{equation}
and the weighted estimates in \eqref{weighted-es}.
Inserting \eqref{toinsert} into the bound for the (localized) Nusselt number, one gets
\begin{equation*}
\Nu\lesssim \delta^2\Ra\left[\Ra^{-\frac 16}\Nu^{\frac 12}+\ln\left(\Ra^{-\frac 13}\Nu \ln^{\frac 12}(\Ra^{\frac 13})\right)\right]+\frac{1}{\delta}\,.
\end{equation*}
Using Young's inequality we are led to
$$\Nu\lesssim \delta^2\Ra\ln(\Nu\Ra^{-\frac 13}\ln^{\frac12}(\Ra^{\frac 13})+\frac{1}{\delta}\,.$$
Choosing
$$\delta\sim (\Ra\ln(\Nu\Ra^{-\frac 13}\ln^{\frac12}(\Ra^{\frac 13})))^{-\frac 13}\,,$$
then 
$$\Nu\lesssim \Ra^{\frac 13}\ln^{\frac 13}(\Nu\Ra^{-\frac 13}\ln^{\frac 12}(\Ra^{\frac 13}))\,,$$
yielding\footnote{Here one uses the fact that $XY\ln^{-\frac 13}(XY)\leq Y$ implies $XY\leq Y\ln^{\frac 13}(Y)$}
\begin{thm}
For solutions of \eqref{RBC}--\eqref{Tbc} with no-slip boundary conditions \eqref{no-slip} the following bound holds:
\begin{equation*}
\Nu\lesssim \Ra^{\frac 13}\ln^{\frac 13}(\ln\Ra) \qquad \mbox{ for } \quad \Pra=\infty\,.
\end{equation*}
\end{thm}
Besides the improvement of the logarithmic correction, the interesting aspect of this result is that the background field method has not been used. That is, this bound relies only on the maximum principle for the temperature and on regularity estimates derived from Stokes equation.
We notice that the authors argue that it is sufficient to establish \eqref{max-reg2}
for \eqref{4order-elliptic} with the free-slip boundary conditions
$$\uz=\partial_z^2\uz=0 \qquad \mbox{ for } z\in \{0,1\}\,,$$
for details see Proof of Lemma 4 in \cite{OS11}.
Passing to this boundary conditions is fundamental  
in order to extend the problem to the whole space by periodicity and therefore being able to use Fourier series techniques in both (horizontal and vertical) variables \footnote{this time the symbol $\hat{\cdot}$ denotes the Fourier transform in $\xx=(\xp,z)$. With $k$ and $\xi$ we denote the horizontal and vertical Fourier variable, respectively.}. 
In Fourier variables, the hessian of $\uz$ can be written as 
$$\widehat{\nabla^2\uz}= 
\begin{bmatrix}k \\\xi \end{bmatrix}\otimes \begin{bmatrix}k \\\xi \end{bmatrix} \frac{|k|^2}{(|k|^2+\xi^2)^2}\hat{\theta}$$
and, from this representation, mimicking a Littlewood-Paley decomposition, one can analyze the small-intermediate and large length scales separately. This ``separation of scales'' is crucial for obtaining the desired bound, since the $L^{\infty}-$norm is critical for Calder\'{o}n-Zygmund type estimates. In

\subsection{Finite Prandtl number}\label{FPNs}
In this section we review some results concerning upper bounds on the Nusselt number at finite Prandtl number. Differently from the previous section, in this case the temperature $T$ and the velocity $\uz$ are no longer instantaneously slaved and the equation for $\uz$ can no longer be written as a fourth order boundary value problem as in the case $\Pra=\infty$ (recall \eqref{4order-elliptic}).
As a consequence, we will see that the application of the background field method becomes somewhat unnatural and new approaches have to be considered. 
In order to present the results concerning the three-dimensional case, we will sometimes implicitly assume regularity of the Navier-Stokes equations. Nevertheless the arguments can be made rigorous by working with Leray weak solutions, i.e. assuming
$$\uu\in L^{\infty}((0,\infty); L^2(\Omega))\qquad \nabla \uu\in L^2((0,\infty); L^2(\Omega))\,.$$

\subsubsection{Background field method approach}
Assuming regularity for the solution of the Navier-Stokes equations,  
 adding the balances \eqref{velocity-balance} and \eqref{pert-balance} we have 
\begin{equation}\label{add}
\frac 12\frac{d}{dt}\left(\|\theta\|_{L^2}^2+\frac{1}{\Pr}\|\uu\|_{L^2}^2\right)=
-\int_{\Omega}\tau'u^z\theta-\|\nabla\theta\|_{L^2}^2+\int_{\Omega}\tau''\theta-\|\nabla \uu\|_{L^2}^2+\Ra\int_{\Omega} T \uz\,,
\end{equation}
Since $\uu$, $T$ (and therefore $\theta$) stay bounded in $L^2$ for all time, 
we may take the long-time limit obtaining
\begin{equation*}
  \lla\int_0^1\tau'\partial_z\theta\, dz\rra=-\lla\int_0^1|\nabla\theta|^2\, dz\rra-\lla\int_0^1(\tau'-a)\theta \uz\, dz\rra-\frac{a}{\Ra}\lla\int_0^1|\nabla \uu|^2\, dz\rra\,,
\end{equation*}
for some $a>0$. Here we used that 
$$\lla \int_0^1 Tu^z\, dz\rra_{\rm{H}}=\lla \int_0^1 \theta u^z\, dz\rra_{\rm{H}}\,,$$
since $\la \uz\ra_{\rm{H}}=0$ by the incompressibility condition.
Combining this identity with \eqref{r-dec}
we obtain the new representation of the Nusselt number
\begin{equation*}
  \Nu=\int_0^1|\tau'|^2\, dz-\lla\int_0^1|\nabla\theta|^2\, dz\rra-2\lla\int_0^1(\tau'-a)\uz\theta\, dz\rra-\frac{2a}{\Ra}\lla\int_0^1|\nabla \uu|^2\, dz\rra\,.
\end{equation*}
Clearly this coincides with \eqref{Nu-rep-bfm} when $a=0$.
If we can select a background profile $\tau$ with $\tau(0)=1, \tau(1)=0$ such that the quadratic form
\begin{equation}\label{quadratic-form}
  \Q\{\theta,\uu\}=\lla\int_0^1|\nabla\theta|^2\, dz+2\int_0^1(\tau'-a)\uz\theta\, dz+\frac{2a}{\Ra}\int_0^1|\nabla \uu|^2\, dz\rra
\end{equation}
is positive definite for all $\theta$ satisfying \eqref{flactuation-equation}, then an upper bound is given by 
\begin{equation*}
  \Nu\leq \int_0^1|\tau'|^2\, dz\,.
\end{equation*}
In \cite{DC96}, Doering and Constantin choose the following background profile
\begin{equation}\label{choice1}
  \tau(z)=
  \begin{cases}
  1-\left(\frac{1}{\delta}-1\right)z & \quad 0\leq z\leq \delta\\
  z & \quad \delta\leq z\leq 1-\delta\\
  \left(\frac{1}{\delta}-1\right)(1-z) & \quad 1-\delta\leq z\leq 1\,.
\end{cases}
\end{equation}
\begin{figure}[h!]
\centering
\includegraphics[scale=0.5]{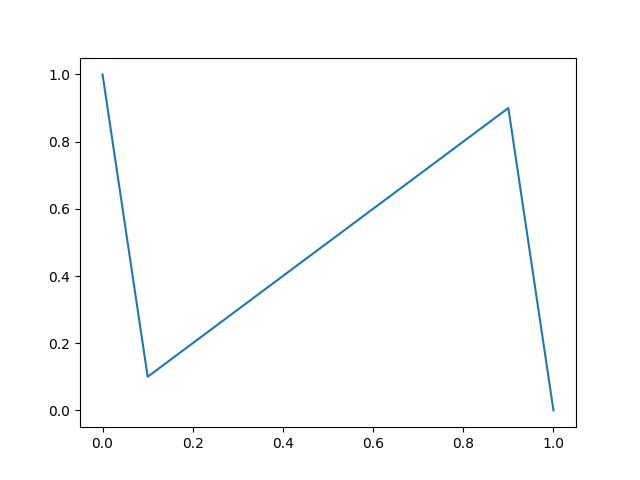}
\caption{Plot of the background profile $\tau=\tau(z)$ used in \cite{DC96}.}
\end{figure}
Inserting this choice in the quadratic form with $a=1$, and focusing on the mixed term we get
\begin{eqnarray*}
  \lla\int_0^1 (\tau'-1)\theta \uz \, dz\rra&=&
  -\frac{1}{\delta}\lla\int_0^{\delta}\theta \uz \, dz+\int_{1-\delta}^1\theta \uz \, dz \rra\,.
\end{eqnarray*}
By Cauchy-Schwarz, Poincar\'e and Young's inequality we obtain 
\begin{eqnarray*}
  \frac{1}{\delta}\lla\int_0^{\delta}\theta \uz \, dz\rra
  &\leq&\delta\lla\int_0^{\delta}|\partial_z \theta|^2\, dz\rra^{\frac 12}\lla\int_0^{\delta}|\partial_z\uz|^2\, dz\rra^{\frac 12}\\
  &\leq&\frac{1}{4}\lla\int_0^{\delta}|\partial_z \theta|^2\, dz\rra+\delta^2\lla\int_0^{\delta}|\partial_z\uz|^2\, dz\rra\,,
\end{eqnarray*}
and the same argument applies to the top boundary layer.
Then  the quadratic form is bounded from below by
$$\Q\{\theta,\uu\}\geq\left(\frac{2}{\Ra}-4\delta^2\right)\lla\int_0^1|\nabla \uu|^2\, dz\rra\,. $$
So, clearly, if $\delta$ is chosen to be
$$\delta\sim \Ra^{-\frac 12}\,,$$
then 
$$\Nu\leq \int_0^1(\tau')^2\, dz\lesssim \frac{1}{\delta}\sim \Ra^{\frac 12}\,.$$
In particular, this shows
\begin{thm}
For solutions of \eqref{RBC}--\eqref{Tbc} with no-slip boundary conditions \eqref{no-slip} it holds
$$\Nu\lesssim \Ra^{\frac 12}\qquad \mbox{ uniformly in } \Pra\,.$$
\end{thm}
 We notice that in this argument neither the incompressibility, nor the full set of boundary conditions was exploited. Therefore this result holds for all boundary conditions such that $\uz=0$. In order to go beyond this result, the system have to be exploited extensively.

\subsubsection{A dynamical systems approach}\label{Wang approach}
In \cite{WA07}, Wang shows that the global attractors of the infinite-Prandtl and the large-Prandtl number convection model remain close, by viewing the Boussinesq system as a small perturbation of the infinite Prandtl number model. Subsequently he derived bounds on the Nusselt number at finite (but large) Prandtl number starting from the Constantin and Doering '99 result in \cite{CD99}.
Following up on his previous works \cite{WA04,WA07,WA08a}, the central idea in Wang's result in \cite{WA08} is to view the Navier-Stokes equations as a perturbation around the stationary Stokes equation,
\begin{equation*}
\begin{array}{rll}
  A\uu:=\nabla p-\Delta \uu&=&f\\
  \nabla \cdot\uu&=&0\,,
\end{array}
\end{equation*}
where $f=\Ra T  \mathbf{e_z}-\frac{1}{\Pra}(\partial_t \uu+\uu\cdot \nabla \uu)$
and apply the bound in \cite{CD99} (reviewed in Section \ref{CD99}).
\newline
Using \eqref{Nu-rep-bfm} and inverting the Stokes operator we have
\begin{equation}\label{Nu-rep-W}
  \Nu= \int_0^1|\tau'|^2\, dz-\lla\int_0^1\, |\nabla \theta|^2+2\Ra\, \tau'\, A^{-1}( \theta\mathbf{e_z})^z\theta-2\frac{1}{\Pra}\tau'A^{-1}(\partial_t\uu)^z\theta -2\frac{1}{\Pra}\tau'A^{-1}((\uu\cdot \nabla)\uu)^z\theta \, dz\rra\,.
\end{equation}
Following the argument in Section \ref{CD99} (applied to the Stokes equation), choosing the background profile \eqref{CD-profile}
then 
$$\lla\int_0^1\, \frac 12|\nabla \theta|^2+2\Ra\, \tau'\, A^{-1}( \theta\mathbf{e_z})^z\theta\, dz\rra\geq 0$$
provided
\begin{equation}\label{delta-choicePr}
  \delta\sim \Ra^{-\frac 13}(\ln \Ra)^{-\frac 23}\,.
\end{equation}
A new argument is needed to estimate the last two terms in \eqref{Nu-rep-W}. In fact, for the first term, using the profile \eqref{CD-profile} together with the Cauchy-Schwarz inequality, Poincar\'e's estimate (twice, in $z$) and Stokes regularity one finds
\begin{eqnarray*}
\left|-2\lla\int_0^1\frac{1}{\Pra}\tau'A^{-1}(\partial_t\uu)^z\theta \, dz\rra\right|
&\leq& 2\delta\frac{1}{\Pr}\lla\int_0^1|\partial_z^2 A^{-1}(\partial_t \uu)^z|^2\, dz\rra^{\frac 12}\lla\int_0^1|\partial_z \theta|^2\, dz \rra\\
&\leq&c\delta^4 \frac{1}{\Pr^2}\lla\int_0^1|\partial_t \uu|^2\, dz\rra^{\frac 12}+\frac 14\lla\int_0^1|\nabla \theta|^2\, dz \rra\,,
\end{eqnarray*} 
for some constant $c=c(\Omega)>0$.
Similarly, using the maximum principle, the bound
\begin{eqnarray*}
\left|-2\frac{1}{\Pr}\lla\int_0^1\tau'A^{-1}((\uu\cdot \nabla)\uu)^z\theta \, dz\rra\right|
&\leq& 2\delta \frac{1}{\Pra}\lla\int_0^1|(\uu\cdot\nabla)\uu|^2\, dz\rra^{\frac 12}\\
&\leq& c\delta \frac{1}{\Pra}\lla\int_0^1|\nabla \uu|^2\, dz\rra^{\frac 34}\lla\int_0^1|A \uu|^2\, dz\rra^{\frac 14}\,,
\end{eqnarray*}
holds, where the spatial $L^2-$norm of the nonlinear term was estimated by the Poincar\'e estimate and Agmon inequality in three dimensions, i.e.
$$\|\uu\cdot\nabla \uu\|_{L^2(\Omega)}\leq\|\uu\|_{L^{\infty}(\Omega)}\|\nabla\uu\|_{L^2(\Omega)}\leq c\|\nabla \uu\|_{L^2(\Omega)}^{\frac 12}\|\Delta \uu\|_{L^2}^{\frac 12} \|\nabla\uu\|_{L^2(\Omega)}\,.$$
The combination of these estimates in \eqref{Nu-rep-W} yields
\begin{eqnarray}\label{final-es}
\Nu\lesssim \int_0^1|\tau'|^2\, dz +\delta^4 \frac{1}{\Pr^2}\lla\int_0^1|\nabla\partial_t \uu|^2\, dz\rra^{\frac 12}+\delta \frac{1}{\Pra}\lla\int_0^1|\nabla \uu|^2\, dz\rra^{\frac 34}\lla\int_0^1|A \uu|^2\, dz\rra^{\frac 14}\,,
\end{eqnarray}
where we used again the Poincar\'e inequality.
Finally, inserting  the choice of $\delta$ in \eqref{delta-choicePr} and the (a-priori) upper bounds
\begin{eqnarray*}
\lla\int_0^1|\nabla \uu|^2\, dz\rra&\lesssim& \Ra^{\frac 32}\\
\lla\int_0^1|A \uu|^2\, dz\rra&\lesssim& \Ra^2\quad \mbox{ for }\; \Pra\gtrsim \Ra\\
\lla\int_0^1|\nabla \partial_t \uu|^2\, dz\rra&\lesssim& \Ra^{\frac 72} \quad \mbox{ for }\; \Pra\gtrsim \Ra\,,
\end{eqnarray*}
in \eqref{final-es}, Wang obtains
\begin{thm}
For solutions of \eqref{RBC}--\eqref{Tbc} with no-slip boundary conditions \eqref{no-slip} the following upper bound holds:
$$\Nu
\lesssim \Ra^{\frac 13}(\ln \Ra)^{\frac 23} \qquad \mbox{ for }\quad \Pr\gtrsim \Ra\,. $$ 
\end{thm}
We remark that the crucial (a-priori) upper bounds used to derive the final estimate on Nusselt number, were obtained by energy and enstrophy inequalities. In particular a Gronwall-type argument is needed to ensure the existence of an absorbing ball such that 
$$\lim_{t\rightarrow\infty}\|\nabla u\|_{L^2}\leq R\,$$
under the large-Prandtl number condition $\Pr\gtrsim \Ra$.

\subsubsection{Maximal regularity approach}
The result in \cite{CNO16} improves Wang's result by perturbing around the nonstationary Stokes equation, i.e. considering 
\begin{equation*}
\begin{array}{rll}
\frac{1}{\Pra}\partial_t \uu+\nabla p-\Delta \uu&=&\Ra T  \mathbf{e_z}-\frac{1}{\Pra}(\uu\cdot \nabla \uu)\\
\nabla \cdot\uu&=&0\,,
\end{array}
\end{equation*}
and proving a \textit{maximal regularity} estimate of the type 
$$\|\nabla^2 \uz\|\lesssim \|\Ra T \mathbf{e_z}-\frac{1}{\Pra}(\uu\cdot\nabla)\uu \|\,,$$
where the norm $\|\cdot\|$ is strong enough to control the nonlinear term
$$\|(\uu\cdot\nabla)\uu\|\lesssim \lla\|\nabla\uu\|_{L^2(0,1)}^2\rra\stackrel{\eqref{diss-bound}}{\lesssim} \Nu\Ra $$
and sufficiently weak so that 
$$\|\Ra T  \mathbf{e_z}\|\lesssim \Ra\lla\|T\|_{L^{\infty}(0,1)}\rra\stackrel{\eqref{MP}}{\lesssim} \Ra\,.$$
The \textit{interpolation norm}, which is able to grant this controls, is
$$\|f\|=\|f\|_{(0,1)}=\sup_{f=f_0+f_1}\left\{ \langle\sup_{0<z<1}|f_0|\rangle+\lla\int_0^1\frac{|f_1|}{z(1-z)}\, dz\rra\right\}\,.$$
In fact, the $L^1$-weighted norm appears natural to bound the nonlinearity:
$$\lla\int_0^1\frac{|(\uu\cdot\nabla)\uu|}{z}\, dz\rra\leq \lla\int_0^1\frac{|\uu|^2}{z^2}\, dz\rra^{\frac 12}\lla\int_0^1|\nabla \uu|^2\, dz\rra^{\frac 12}\leq \lla\int_0^1|\nabla \uu|^2\, dz\rra\,,$$
where we used the Cauchy-Schwarz inequality together with Hardy's inequality and the dissipation bound in \eqref{diss-bound}.
Since the interpolation norm is critical for maximal regularity estimates (since the $L^{\infty}$ and the $L^1-$weighted norm are both critical with this respect), the authors proved the following
\begin{prop}
Suppose $\uu,p$ and $\mathbf{f}$ satisfy
\begin{equation*}
\begin{array}{rlll}
\frac{1}{\Pra}\partial_t \uu+\nabla p-\Delta \uu&=&\mathbf{f} & \mbox{ in }z\in(0,1)\\
\nabla \cdot\uu&=&0  & \mbox{ in }z\in(0,1)\\
\uu&=&0 &\mbox{ at }z=\{0,1\}\\
\uu&=&0 & \mbox{ at }t=0\,,
\end{array}
\end{equation*}
and $\mathbf{f}$ is horizontally banded, i.e.
$$\hat {\mathbf{f}}(k,z,t)=0\quad \mbox{ unless }\;  1<R|k|<4\,.$$
Then
\begin{equation}\label{MR}
\|(\partial_t-\partial_z^2)\up\|+\|\partial_t\uz\|+\|\nabla \nabla'\up\|+\|\partial_z^2\uz\|+\|\nabla p\|\lesssim \|\mathbf{f}\|\,,
\end{equation}
holds.
\end{prop}
As we can see, the authors do not obtain the full maximal regularity estimate due to cancellations that occurs as an effect of using the bandedness assumption. Nevertheless the crucial estimate
$$\|\partial_z^2\uz\|\lesssim \|f\|$$
is not affected and allows to conclude the argument.
In fact, if only horizontal intermediate wavelengths could be taken into account, then, inserting the maximal regularity estimate in the bound
$$\Nu
  \stackrel{\eqref{av-Nusselt}}{\lesssim} \delta^2\|\partial_z^2\uz\|+\frac{1}{\delta}\,, 
$$
and choosing $\delta\sim ((\frac{\Nu}{\Pra}+1)\Ra)^{-\frac 13}$, we would be able to conclude
$$\Nu\lesssim\begin{cases}
 (\Ra)^{\frac 13}& \Pra\gtrsim(\Ra)^{\frac 13}\\
(\frac{\Ra}{\Pra})^{\frac 12} &  \Pra\lesssim(\Ra)^{\frac 13}\,.
\end{cases}$$

This estimate is unfortunately not achieved as the high and low wavelength will weight in the final estimate. Nevertheless, by the incompressibility condition and the a-priori energy inequality \eqref{diss-bound}, the authors show that the region of intermediate wavelengths (where estimate \eqref{MR} holds) is large and that the correction from the desired bound is relatively small. In conclusion, we have
\begin{thm}
For solutions of \eqref{RBC}--\eqref{Tbc} with no-slip boundary conditions \eqref{no-slip}, the following upper bound holds:
$$\Nu\lesssim
 \begin{cases}
 (\Ra\ln\Ra)^{\frac 13}& \Pra\gtrsim(\Ra\ln\Ra)^{\frac 13}\\
(\frac{\Ra}{\Pra}\ln\Ra)^{\frac 12} &  \Pra\lesssim(\Ra\ln\Ra)^{\frac 13}\,.
\end{cases}
$$
\end{thm}
\section{Free-slip and Navier-slip boundary conditions}\label{four}
In the previous sections, the no-slip boundary conditions 
$$\uz=0 \quad \mbox{ and }\quad \partial_z\uz=0 \quad \mbox{ at } z=\{0,1\}$$
\footnote{Recall from Section \ref{ipns} that this is an equivalent way of writing the no-slip boundary conditions using incompressibility.}
have been crucial to obtain upper bounds of the type $\Nu\lesssim (\log\Ra)^{\gamma}\Ra^{\frac 13}$.
In fact by the application of Poincar\'e's estimate in $z$ we bounded
$$\|u\|_{L^p(0,\delta)}\leq\delta^2\|\partial_z^2u\|_{L^p(0,\delta)}\,.$$
With free-slip and Navier-slip boundary conditions this estimate is no longer available and it is therefore more difficult to appeal to maximal regularity estimates, which turned out to be crucial for producing optimal bounds in the no-slip setting.
In the next section we see how to circumvent this problem and get sharp quantitative estimates on the flow. We want to notice here that, in terms of regularity estimates, the free-slip boundary conditions present a big advantage compared to the no-slip boundary conditions. In fact the free-slip boundary conditions \eqref{free-slip} are equivalent to
$$\uz=0 \quad \mbox{ and }\quad \partial_z^2\uz=0 \quad \mbox{ at } z=\{0,1\}$$
since $\partial_z^2\uz=-\nabla_H\cdot\partial_z\up=0$ thanks to the incompressibility condition.
Now one can easily see that the equations can be naturally extended in the vertical direction by periodicity. This observation was used in \cite{OS11} to derive a maximal regularity estimate and in \cite{CJTW21} to show the existence of a global attractor in each affine space where the velocity has fixed spatial
average. One major advantage is that the extension by periodicity allows to use Fourier-series techniques also in the vertical direction.  In the no-slip case, this extension was not possible, and, in order to derive maximal regularity bounds we were forced to decompose the operator and exploit Fourier-series techniques only in the horizontal variables.

Due to limitations in the analysis at finite Prandtl number, in the next sections we will often consider the two dimensional Rayleigh-B\'enard system \eqref{RBC}. In this case we will denote with $\xx=(x,z)$ a vector in $\Omega=[0,L]\times[0,1]$ and with $\uu=(\ux,\uz)$ the velocity field.
\subsection{Free-slip boundary conditions}
\subsubsection{Bounds in two dimensions and at finite Prandtl number}
In his PhD thesis in 2002 \cite{O02}, Jesse Otero reported numerical indication of the scaling $\Nu\sim \Ra^{\frac{5}{12}} $ for the two-dimensional Rayleigh-B\'enard convection problem with free slip boundary conditions. 
Later in 2011, Doering and Whitehead in \cite{WD11} proved rigorously the following result
\begin{thm}\label{2d-freeslip}
For solutions of \eqref{RBC}--\eqref{Tbc} and \eqref{free-slip} in two-dimensions the following bound holds:
\begin{equation}\label{512}
\Nu\lesssim \Ra^{\frac{5}{12}} \qquad \mbox{ uniformly in }\; \Pra.
\end{equation}
\end{thm}
The argument takes advantage of information on the scalar vorticity $\omega=\partial_x\uz-\partial_z\ux$ satisfying 
\begin{equation}\label{vort-eq}
\begin{array}{rrll}
\frac{1}{\Pra}\left(\partial_t\omega+\uu\cdot\nabla \omega\right)-\Delta \omega &=&\Ra \partial_xT & \mbox{ in } \Omega=[0,L]\times [0,1]\\
\omega&=&0 & \mbox{ at } z=\{0,1\}\,.
\end{array}
\end{equation}
The enstrophy balance is
\begin{equation}\label{enst-bal}
\frac{1}{2\Pr}\frac{d}{dt}\|\omega\|_{L^2(\Omega)}^2=-\|\nabla \omega\|_{L^2(\Omega)}^2+\Ra\iint_{\Omega} \omega \partial_x\theta dx\, dz\,.
\end{equation}
from which one can show \cite{CJTW21} that 
$$\|\omega\|_{L^2(\Omega)}\leq C \Ra$$
 and obtain
$$0=-\lla\|\nabla \omega\|_{L^2}^2\rra+\Ra\lla\int_0^1 \omega \partial_x\theta\, dz\rra\,.$$
Combining this, with the long-time averages of \eqref{pert-balance} and \eqref{velocity-balance}, the authors obtain the representation
$$\Nu=\frac{1}{1-b}\left(\int_0^1|\tau'(z)|^2\, dz-b\right)-\frac{1}{1-b}\Q $$
where 
\begin{equation}\label{quad-form-fs}
\Q=\lla\int_0^1\left(|\nabla \theta|^2+\frac{a}{\Ra^{\frac 32}}|\nabla \omega|^2+\frac{b}{\Ra}|\omega|^2+2\tau'\uz\theta+\frac{a}{\Ra^{\frac12}}\omega \partial_x\theta\,\right) dz\rra\,,
\end{equation}
with $b\in (0,1)$ and $a>0$.
By standard arguments it is not difficult to show that 
$$\hat{\Q}\geq\|\partial_z\hat{\theta}\|_{L^2}^2+\left[\frac{ak^2}{\Ra^{\frac 32}}+\frac{1}{\Ra}\left(b^2-\frac{a^2}{4}\right)\right]\|\hat{\omega}\|_{L^2}^2-\frac{1}{\delta}\Real\left\{\int_0^{\delta}\hat{\uz}\hat{\theta}^{\ast}\, dz+\int_{1-\delta}^{1}\hat{\uz}\hat{\theta}^{\ast}\, dz\right\}\,,$$
where $\Real$ denotes the real part of a complex number.
Since the first two terms of the right-hand side are already non-negative (for some choice of $a$ and $b$), the authors need to argue for the (potentially negative) mixed term. 
%
We start with an observation: 
 since 
$$\int_0^1 \partial_z \hat\uz\, dz=\hat\uz(1)-\hat\uz(0)=0\,,$$
there exists a $z_0\in (0,1)$ (no matters where it lies in $(0,1)$!) such that 
$$\partial_z \hat\uz(z_0)=0$$ 
Then the ``missing'' derivative (for applying Poincar\'{e}'s estimate) is partially recovered by the estimate
\begin{eqnarray*}
 |\partial_z \hat\uz(z)|^2 &=& \int_{z_0}^{z}\partial_z (\partial_z \hat\uz(z'))^2 \, dz'\\
 &=&2\int_{z_0}^{z} \partial_z \hat\uz(z')\partial_z^2 \hat\uz(z')\, dz'\\
 &\leq&\frac{1}{k}[2c_1 \| \partial_z \hat\uz\|^2+\frac{k^2}{2c_1}\|\partial_z^2 \hat\uz\|^2]\,,
 \end{eqnarray*}
 where we used the Youngs inequality.
Then 
$$|\hat\uz(z)|=\left|\int_0^z\partial_z \hat\uz(z')\, dz'\right|\leq z\left(\int_0^1|\partial_z \hat\uz|^2\, dz\right)^{\frac 12}
\leq \frac{z}{\sqrt k}[2c_1\|\partial_z \hat\uz\|_{L^2}^2+\frac{k^2}{2c_1}\|\partial_z^2 \hat\uz\|_{L^2}^2]^{\frac 12}$$
Combining this estimate with 
 $$|\hat\theta(z)|\lesssim z^{\frac 12}\|\partial_z \hat\theta(z)\|_2\,,$$
 which holds thanks to the assumptions on $\theta$, we showed
 \begin{lemma}\label{crucial}
For $w\in H^2(0,1)$ with $w=0$ at $z=\{0,1\}$ and $\zeta\in H^1(0,1)$ with $\zeta=0$ at $z=\{0,1\}$, the bound
\begin{equation}\label{b-512}
\frac{1}{\delta}\int_0^{\delta}w(z)\zeta^{\ast}(z)\, dz
\leq\frac{2}{25k}\delta^3\left[2c_1\|\partial_z^2 w\|_{L^2}^2+\frac{k^2}{2c_1}\|\partial_zw\|_{L^2}^2\right]+\frac 12\|\partial_z \zeta\|_{L^2}^2\,,
\end{equation}
holds for some positive constants $c_1$.
\end{lemma}
We observe now that by using the (2d) relation $\omega=\partial_z\ux-\partial_x\uz$ and the divergence-free condition in Fourier variables, one obtains
$$ik \hat\omega=\partial_z^2\hat \uz-k^2\hat\uz\,,$$
and easily find the bound
\begin{equation}\label{vort-rel}
\frac{8}{9}\|\partial_z^2\hat \uz\|_{L^2}^2+\frac 83k^2\|\partial_z\uz\|_{L^2}^2\leq k^2\|\hat{\omega}\|_{L^2}^2\,.
\end{equation} 
In particular, \eqref{b-512} and \eqref{vort-rel} together with the choice $c_1=\frac{\sqrt{3}}{6}$ yield
\begin{equation}\label{est-uz-DW}
\quad|\hat \uz(z)|\lesssim k^{\frac 12}z\|\hat\omega(z)\|_{L^2}\,.
\end{equation}
Combining this a-priori bound with Lemma \ref{crucial} applied to $w=\hat{\uz}$ and $\zeta=\hat{\theta}$, we find
\begin{equation}\label{crucial-512}
\frac{1}{\delta}\int_0^{\delta}\hat{\uz}(z)\hat{\theta}^{\ast}(z)\, dz
\leq c \delta^3k\|\hat{\omega}\|_{L^2}^2+\frac 12\|\partial_z \hat\theta\|_{L^2}^2\,.
\end{equation}
Inserting \eqref{crucial-512} in the lower bound for $\hat{\Q}$, we immediately see that $\hat{\Q}\geq 0$ if
$$\frac{ak^2}{\Ra^{\frac 32}}+\frac{1}{\Ra}\left(b-\frac{a^2}{4}\right)-c\delta^3k\geq 0\,.$$
Minimizing on $\delta$ (over $k$), choosing $\delta\sim \Ra^{-\frac{5}{12}}$ where
$k\sim \Ra^{\frac 14}$ is the minimal wavenumber, we obtain the desired bound $\Nu\lesssim \Ra^{\frac{5}{12}}$. 

Recent numerical simulations \cite{WGLCD20} indicate the clear asymptotic scaling
$$\Nu\sim\Ra^{\frac 13}\qquad\qquad \mbox{for} \quad \Ra\rightarrow \infty$$
for steady $2d$ convection rolls between free-slip boundary conditions. It is an interesting problem to bridge this gap and see whether the result in \cite{WD11} is optimal or steady $2d$ convection rolls (slightly) inhibit heat transport.

\subsubsection{Bounds in three dimensions}
In three dimensions, the vorticity is no longer a scalar and the enstrophy balance \eqref{enst-bal} does not hold. Then the variational problem to study is: find the optimal $\tau$ that minimizes the right-hand side of 
$$\Nu\leq\frac{1}{1-b}\left(\int_0^1|\tau'|^2\, dz-b\right)$$
constrained to
$$\Q\{\theta\}=\lla\int_0^1|\nabla\theta|^2+\frac{b}{\Ra}|\nabla \uu|^2+2\tau'\uz\theta\, dz \rra\geq 0\,.$$
This is the same as \eqref{quad-form-fs} with $a=0$. 
In \cite{PI05} and \cite{IKP06} the authors apply the background field method to the free-slip problem and their numerical simulations indicate 
$$\Nu\sim 0.1262\Ra^{\frac{5}{12}}\,$$
for stably stratified profiles in the bulk, with $\tau'=p>0$. Choosing this family of background profiles, 
Doering and Whitehead in \cite{WD12} rigorously proved 
\begin{thm}
For solutions of \eqref{RBC}--\eqref{Tbc} and \eqref{free-slip} in three dimensions the following upper bound holds:
\begin{equation}\label{opt-const}
\Nu\lesssim\Ra^{\frac{5}{12}} \qquad \mbox{ with }\quad \Pra=\infty\,.
\end{equation}
\end{thm}
 confirming the optimality of the family of stably stratified profiles and the numerical result in \cite{PI05}. The argument is almost identical to the one for Theorem \ref{2d-freeslip}, with the crucial bound \eqref{est-uz-DW} replaced by 
\begin{equation}\label{es-pseudo-vort}
 |\hat{\uz}(z)|\lesssim k^{\frac 12}z\|\hat{\zeta}\|_{L^2}\,,
\end{equation}
where $\zeta$ imitates the role of vorticity in two dimensions and is referred to as ``pseudo-vorticity'' (or ``pseudo-enstrophy'') and it is defined by
\footnote{In real space the pseudo vorticity is defined through the relation $$\Delta \uz=|\nabla_{\rm H}|\zeta$$} 
\begin{equation}\label{pseudo-vor}
(\partial_z^2-k^2)\hat{\uz}=k\hat{\zeta}\,.
\end{equation}
Recalling the elliptic fourth-order boundary value problem written in Fourier variables \eqref{4order-FT}, the pseudo-vorticity satisfies
\begin{eqnarray}\label{relat-pseudo-vor-T}
(\partial_z^2-k^2)\hat{\zeta}&=&\Ra k\hat{T}\qquad \mbox{in } \Omega\\
\hat{\zeta}&=&0 \qquad\qquad \mbox{at } z\in \{0,1\}\,.
\end{eqnarray}
An interesting point  discussed in \cite{WD12} concerns the relation between the slope $p$ of the profile $\tau$ and the optimal constant $c$ in \eqref{opt-const}. Fist, the authors clarify that, a stratified profile is the bulk or a balance parameter $b$ with a non-stratified profile, produce the same scaling. In fact the authors demonstrate how setting $p=0$ and optimizing in $b$ produce a prefactor $c$ that is almost indistinguishable from the case in which $b=0$ and $p$ is optimized. In particular, setting $b=0$ and $p=\frac{3}{29}$, the authors find $\Nu\lesssim0.2982\Ra^{\frac{5}{12}}$. As the slope parameter is nearly the same to $p\sim 0.103$ computed by Plasting and Ierley in 2005, the analysis of Whitehead and Doering is rather sharp.  Second, the author showed that the best constant ($0.28764$) is achieved for $p=-\frac{1}{17}$ and $b=\frac{5}{12}$.  This in particular tells that, if $b>0$, the (optimal) bound is produced with a non-stably stratified profile (since the slope parameter is negative).

\medskip

Combining the methods in \cite{WA08} and \cite{WD12}, Wang and Whitehead in \cite{WW13} succeeded in treating the three-dimensional finite Prandtl number case as a perturbation of the infinite Prandtl number problem problem, i.e. studying
\begin{equation*}
\begin{array}{rll}
A\uu:=\nabla p-\Delta \uu&=&\Ra T  \mathbf{e_z}-\frac{1}{\Pra}\f\\
\nabla \cdot\uu&=&0\,,
\end{array}
\end{equation*}
where $\f=\partial_t \uu+\uu\cdot \nabla \uu$ and $\Pra$ is supposed to be (very) large, allowing them to apply the argument in \cite{WD12}. In particular, the authors eliminate the pressure and study the relation
$$\Delta^2\uz=-\Ra\Delta_{\rm H}T+\frac{1}{\Pr}\left(-\DeltaH\fz+\partial^2_{xz}\fx+\partial^2_{yz}\fy\right)\,,$$
and benefit from rewriting the equation in Fourier space 
$$k(-k^2+\partial_z^2)\hat{\zeta}=\Ra k^2\hat{T}+\frac{1}{\Pra}(k^2\hat{\fz}+ik_1\partial_z\hat{\fx}+ik_2\partial_z\hat{\fy})\,,$$
where $\zeta$ is the pseudo-vorticity defined in \eqref{es-pseudo-vort}. Modulo the $O(\frac{1}{\Pra})$ inertial terms, this relationship is identical to \eqref{relat-pseudo-vor-T}.

 Also here, as in Section \ref{Wang approach}, the idea is to use the background field method, splitting the quadratic form
$$\Q\{\theta\}=\lla\int_0^1(|\nabla \theta|^2+2\tau'\uz\theta)\, dz\rra\,.$$
into a part (containing $O(1)$ contributions already present at infinite Prandtl number) which is positive definite and a rest, of order $o(\frac{1}{\Pr})$ (and potentially negative) that needs to be controlled with other methods.
They proved
\begin{thm}
Suppose that $\uu, T$ solve \eqref{RBC}--\eqref{Tbc} and \eqref{free-slip}. Then there exists a non-dimensional constant $c_0$ such that if the Grashof number $\Gr$ is sufficiently small, i.e,
$$\Gr=\frac{\Ra}{\Pra}\leq c_0\,,$$
then
$$\Nu\lesssim \Ra^{\frac{5}{12}}+\Gr^2\Ra^{\frac 14}\,. $$
\end{thm}
Here $c_0$ is a small number determined by ensuring the simultaneous validity of 
$$\lla\int_0^1 |A\uu|^2\, dz\rra\lesssim \Ra^2$$ 
$$\lla\int_0^1 |\nabla \partial_t\uu|^2\, dz\rra\lesssim \Ra^{\frac 72}\, \qquad \mbox{and}$$
$$\lla\int_0^1 |\nabla((\uu\cdot \nabla)\uu)|^2\, dz\rra\lesssim\left(\Ra^{\frac{17}{4}}+\frac{1}{\Pra^{\frac 32}}\Ra^5+\frac{1}{\Pra^{6}}\Ra^{\frac{19}{2}}\right)\,.$$
These are the fundamental bounds necessary to control the $O(\frac{1}{\Pr})$ negative contributions in the quadratic form.

\subsection{Navier-slip boundary conditions}
As observed before, while free-slip conditions are considered to be ``non-physical'' due to the absence of vorticity production at the boundary (i.e. $\omega=0$ at $z=\{0,1\}$), the
Navier-slip conditions are adapt to describe surfaces where some slip occurs, but there is some stress on the fluid. In fact, in realty most of the surfaces are neither perfectly free slip not no slip \cite{WD12}. If we think of this boundary conditions as interpolating between no-slip and free-slip (as the slip length $\LL$ increases from $0$ to $\infty$), we expect that the bounds on Nusselt number will interpolate as well, between the $\Ra^{\frac 12}$ and the $\Ra^{\frac{5}{12}}$ bounds.
In two dimensions, the vorticity satisfies
\begin{equation}\label{vort-freeslip}
\begin{array}{rrll}
\frac{1}{\Pra}\left(\partial_t\omega+\uu\cdot\nabla \omega\right)-\Delta \omega &=&\Ra \partial_x T & \mbox{ in } \Omega=[0,L]\times[0,1]\\
-\omega &=&\frac{1}{\LL}\ux & \mbox{ at } z=0\\
\omega &=&\frac{1}{\LL}\ux & \mbox{ at } z=1\,,
\end{array}
\end{equation}
and the enstrophy balance is 
\begin{eqnarray}\label{en-bal}
&&\frac{1}{2\Pra}\frac{d}{dt}\|\omega\|_{L^2(\Omega)}^2+\frac{1}{2\LL\Pra}\frac{d}{dt}\left(\|\ux\|_{L^2(\{z=1\})}^2-\|\ux\|_{L^2(\{z=0\})}^2\right)+\|\nabla \omega\|_{L^2(\Omega)}^2\notag\\
&&=\frac{1}{\LL}\left(\int_0^{L}p\partial_{x}\ux|_{z=1}\, dx+\int_0^{L}p\partial_{x}\ux|_{z=0}\, dx\right)+\Ra\int_{\Omega}\omega \partial_xT\, dx\, dz\,.
\end{eqnarray}
Notice that as $\LL\rightarrow \infty$, this balance reduces to \eqref{enst-bal}.
Following an idea in \cite{LLP05} applied to the two-dimensional unforced Navier-Stokes equations the authors prove
\begin{lemma}
Let $\omega$ satisfy \eqref{vort-freeslip}, $\LL \geq 1$ and $p \in [1, \infty)$. Then 
$$\|\omega(t)\|_{L^p}\lesssim\|\omega_0\|_{L^p}+\frac{1}{\LL}\|u_0\|_{L^2}+\Ra\,.$$
\end{lemma}
Thanks to this lemma, combining the averaged version on \eqref{en-bal} with the balance
$$\lla\int_0^1|\nabla u|^2\, dz\rra+\frac{1}{\LL}\left(\lla(\ux)^2|_{z=1}\rra+\lla(\ux)^2|_{z=0}\rra\right)-\Ra(\Nu-1)=0\,,$$
we can write a new representation of the Nusselt number:
$$(1-b)\Nu+b=\int_0^1|\tau'|\,dz+M\Ra^2-\Q\{\theta,\uu\}$$
where 
\begin{eqnarray*}
\Q\{\theta,\uu\}
 &:=&M\Ra^2+\lla\int_0^1 |\nabla \theta|^2\, dz\rra+\frac{b}{\Ra}\lla\int_0^1|\omega|^2\, dz\rra+a\lla\int_0^1|\nabla \omega|^2\, dz\rra\\
 &+&2\lla\int_0^1\tau'\uz\theta\, dz\rra+\frac{b}{\Ra \LL}\left(\lla (\up)^2|_{z=1} \rra+\lla (\up)^2|_{z=0} \rra\right)\\
 &-&\frac{a}{\LL}\left(\lla p\partial_{x}\ux|_{z=1}\rra+\lla p \partial_{x}\ux|_{z=0}\rra\right)
 -a\Ra\lla\int_0^1\omega\partial_{x}\theta\, dz\rra,
\end{eqnarray*}
with parameters $a>0$, $b\in(0,1)$ and $M>0$ to be chosen at the end.
The combination of the arguments in \cite{WD11} (to bound the mixed term $\la\int_0^1\tau'\uz\theta\, dz\ra$), Sobolev trace inequalities together with the regularity estimate on the pressure contained in
\begin{prop}
The pressure in \eqref{RBC} satisfies 
\begin{equation*}
\begin{array}{rrll}
\Delta p &=&-\frac{1}{\Pra}\nabla \uu^t:\nabla u+\Ra\partial_zT & \mbox{ in } \Omega\,,\\
-\partial_yp &=&\frac{1}{\LL}\partial_x\ux-\Ra & \mbox{ at } z=0\,,\\
\partial_yp &=&\frac{1}{\LL}\partial_x\ux & \mbox{ at } z=1\,,
\end{array}
\end{equation*}
and for any $r\in (2,\infty)$ the estimate
$$\|p\|_{H^1}\lesssim \frac{1}{\LL}\|\partial_x\omega\|_{L^2}+\Ra\|T\|_{L^2}+\frac{1}{\Pra}\|\omega\|_{L^2}\|\omega\|_{L^r}$$ 
holds\,.
\end{prop} 
 yield the lower bound for $\Q$
\begin{eqnarray*}
   \Q\{\theta,u\}
   &\geq& \frac 12\lla\int_0^1|\nabla  \theta|^2\, dz\rra
   +a\left(\frac 14-\frac{C}{\LL^2}\right)\lla\int_0^1|\nabla\omega|^2\, dz\rra\\
   &+&\left(\frac{b}{\Ra}-\frac{a^2\Ra^2}{2}-\frac{aC^2\|u_0\|_{W^{1,r}}}{2\LL^2\Pra^2}-\frac{aC^2\Ra^2}{2\LL^2\Pra^2}-c\delta^6a^{-1}\right)\lla\int_0^1|\omega|^2\, dz\rra\,.
  \end{eqnarray*} 
  Optimizing in $a,b,M$ and choosing $\delta\sim \Ra^{-\frac{5}{12}}$ the authors find
 \begin{thm}
 Let $u,T$ solve \eqref{RBC}--\eqref{Tbc} with Navier-slip boundary conditions \eqref{Navier-slip}. Then
\begin{equation}\label{result-NS}
\Nu\lesssim \Ra^{\frac{5}{12}}+\LL^{-2}\Ra^{\frac 12} \qquad \mbox{ for }\quad \LL^2\Pra^2\geq \Ra^{\frac 32}\,.
\end{equation}
\end{thm}
 Moreover, for the case $\Pra=\infty$ in dimension $d\geq 2$, the authors show that the bound 
$$\Nu\lesssim \Ra^{\frac{5}{12}}\,$$
follows from the same arguments. Using similar techniques to the one in \cite{WD12}, this upper bound was independently (and previously) noticed by Whitehead in unpublished notes.

\section{Conclusion and open problems}

In this paper we reviewed a selection of contributions in the field of quantitative bounds on the Nusselt number for the classical Rayleigh-B\'enard convection problem between fixed impermeable horizontal boundaries. In the following chart we summarize the best upper bounds available, distinguishing between no-slip, Navier-slip and free-slip boundary conditions.

\vspace{1cm}
{\small
\begin{tabular}{ |p{2cm}||p{4.5cm}|p{4.5cm}|p{4.5cm}|  }
 \hline
 \multicolumn{4}{|c|}{Summary of best known results} \\
 \hline
 \textbf{Boundary conditions}& \textbf{No-slip} & \textbf{Navier-Slip} & \textbf{Free-slip}\\
 \hline
 \textbf{2d,3d}   \newline  $\Pra=\infty$   & $\Nu\lesssim (\ln\ln \Ra)\Ra^{\frac13}$    &$\Nu\lesssim \Ra^{\frac{5}{12}}$ &   $\Nu\lesssim \Ra^{\frac{5}{12}}$\\
 & \small{(Otto\&Seis '15)}& \small{(Whitehead/\qquad\qquad  \qquad \qquad Drivas, Nguyen, Nobili '21)}& \small{(Whitehead\&Doering '11)}\\
 \hline
 
 \textbf{2d}, $\Pra<\infty$ & \small{$\Nu\lesssim \begin{cases}(\ln\Ra\Ra)^{\frac 13} & \Pra\gtrsim \Ra^{\frac 13}\\ (\ln\Ra\frac{\Ra}{\Pr})^{\frac 12} & \Pra\lesssim \Ra^{\frac 13}\end{cases}$} \newline $\Nu\lesssim \Ra^{\frac{1}{2}}$ &   \small{$\Nu\lesssim \Ra^{\frac{5}{12}}+\LL^{-2}\Ra^{\frac 12}$ } \newline \footnotesize{ for $\LL^2\Pr^2\geq \Ra^{\frac 32}$}&  $\Nu\lesssim \Ra^{\frac{5}{12}}$\\
 &\small{(Choffrut, Nobili, Otto '16, \newline Doering\&Constantin'96)} & \small{(Drivas, Nguyen, Nobili '21)} & \small{(Whitehead\&Doering '11)}\\
 \hline
 \textbf{3d}, $\Pra<\infty$    & same as 2D for Leray-weak solutions & not known&$\Nu\lesssim \Ra^{\frac{5}{12}}+\rm{Gr}^2\Ra^{\frac 14}$ \newline \footnotesize{for $\rm{Gr}$ sufficiently small}  \\
 &  & &\small{(Wang\&Whitehead '12)}\\
 \hline
\end{tabular}}
\vspace{1cm}

\noindent
There are various questions and open problems that remain open:

\textit{Ultimate scaling: $\Ra^{\frac 12}$ or $\Ra^{\frac 13}$?}
 Focusing on the no-slip boundary conditions, we see that, on the one hand, upper bounds of the type $\Nu\lesssim \Ra^{\frac 13}(\log\Ra)^{\gamma}$ hold only for $\Pra=\infty$ or for very large Prandtl number (i.e. $\Pra\gtrsim (\Ra\ln\Ra)^{\frac 13}$). On the other hand, the upper bound $\Nu\lesssim\Ra^{\frac 12}$ holds uniformly in $\Pra$ and it improves the bound $\Nu\lesssim (\Ra\ln\Ra\Pra^{-1})^{\frac 12}$ for low Prandtl numbers. At the same time,  this bound is the same no matter what the boundary conditions are, as long as $\uz=0$ at $z=\{0,1\}$. Thus the analysis is not yet able to confirm or rule out any of the scaling theories. Also, direct numerical simulations see their limitations in power of prediction when $\Ra$ becomes very big and often the data points reported are not sufficient to conclude a scaling behavior as argued in \cite{DTW19,D20}. Nevertheless, in \cite{ISSS20} the authors show that the three dimensional simulations-data fit $\Nu\sim \Ra^{0.331}$ in a convection cell with small aspect ratio, $\Pr= 1$ and $\Ra\in[10^{10},10^{15}]$.

\textit{Do boundary conditions matter in the ultimate regime?}
  Whether the boundary conditions matter in the ultimate--turbulent--state, remains to be understood. Nevertheless the results in the chart seem to indicate that free-slip enhance vertical heat transport compared to no-slip boundary conditions. On the other hand, for the choice $\LL\sim\Ra^{\alpha}$ and $\Pr\geq \Ra^{\frac 34-\alpha}$ the bound in \eqref{result-NS} reads 
$$ \Nu\lesssim
\begin{cases}
\Ra^{\frac{5}{12}} & if \quad\alpha\geq \frac{1}{24}\\ 
\Ra^{\frac 12-2\alpha} & if \quad 0\leq\alpha\leq \frac{1}{24}\,.
\end{cases} 
$$
If we compare this bound for small slip-lengths with the upper bound for no-slip boundary conditions at relatively small Prandtl numbers ($\Nu\lesssim \Ra^{\frac 12}$) we observe that the Navier-slip conditions seem to slightly inhibit heat transport.

\textit{Dimensions: can we say something more in $2d$?}
Again looking at the chart we can see that little difference can be appreciated between the results in two and three dimensions, especially for no-slip boundary conditions. 
While the analysis in three dimensions is limited by the complexity of the Navier-Stokes equations, the two dimensional dynamics are accessible and might reveal significant mechanism in heat transport. 
In fact, the two dimensional Rayleigh-B\'enard problem displays many of the physical and turbulent transport feature of three dimensional convection and the structure and regularity of the flow allows finer estimates. In fact, as seen in \cite{WD12,DNN21} balances involving the vorticity (and its gradient) can be supplemented to the energy balance to discover important relations. The next challenge in mathematical analysis would be to improve the upper bounds in \cite{CNO16}, as the arguments in this paper are blind towards dimensions. In particular, in two dimensions, can one extend the regime of validity of the bound $\Nu\lesssim \Ra^{\frac 13}$ at finite Prandtl number? or can one prove a bound of the type $\Nu\lesssim \Ra^{\beta}$ with $\beta<\frac 13$ in some (possibly low) Prandtl number regime? 

\textit{Are these results optimal?}
Let us notice that all rigorous results are in the form of \textit{upper bounds}, which therefore give us indications about the ``worst case'' scenario. It remains to be seen whether these estimates are sharp. 
This is very challenging from a mathematical standpoint and, at the current time, no lower bounds for the Nusselt number have been derived. Constructing exact solutions that saturate the bounds is probably impossible, given the complexity of the system, even at infinite Prandtl number. Nevertheless, in this direction, we want to mention the works \cite{TD17,DT19} in which Tobasco and Doering consider a passive tracer advected by an incompressible velocity field $\uu$, satisfying $\la|\nabla \uu|^2\ra\leq \rm{Pe}^2$, where $\rm{Pe}$ is the Peclet number. By adopting optimal designing techniques in energy-driven pattern formation in material science, 
the authors succeeded in designing an incompressible flow satisfying the enstrophy constraint such that $\Nu\gtrsim \rm{Pe}^{\frac 23}(\log\rm{Pe})^{-\frac 43}$. Together with the upper bound $\Nu\lesssim \rm{Pe}^{\frac 23}$ \cite{HCD14}, this lower bound demonstrate $\Nu\sim \rm{Pe}^{\frac 23}$ possibly up to a logarithm. In connection with the Rayleigh-B\'enard convection problem, this result shows the existence of a flow which is not buoyancy driven but nevertheless realizes $\Nu\sim \Ra^{\frac 12}$.

\textit{What are the bounds in case of rough boundaries?}
Another important problem to study is the role of rough boundaries in the bounds. Numerical studies \cite{SW15} indicate that roughness can enhance heat transport.
To the author's knowledge, the only rigorous result available is given by Goluskin and Doering in 2016. In their paper \cite{GD16}, they assume no-slip boundary conditions on a rough vertical walls. More precisely, the (vertical) boundaries are continuous (and piecewise differentiable functions) of the horizontal coordinates (that is $z=h^T(x,y)$ and $z=h^B(x,y)$) with finite $L^2$ gradients. By a (not standard) application of the background field method, the authors show
$$\Nu\lesssim C(\|\nabla h\|_{L^2}^2)\,\Ra^{\frac 12}\,.$$
The scaling exponent $\frac 12$ was observed in the experiments reported in \cite{RCCH01} where all boundaries (including sidewalls) were rough and, most recently, in numerical simulations using sinusoidally rough upper and lower surfaces in two dimensions \cite{TSW17}. Nevertheless the bound derived by Goluskin and Doering depends on a constant of the $L^2$-norm of the gradient of $h$. It is expected that this condition can be relaxed, allowing much weaker regularity. 
As remarked in the introduction, it is not clear whether the no-slip boundary conditions are the most suitable to describe the physical problem. 
Citing \cite{ZG02} ``\textit{For many years it has been observed that there is no
compelling argument to justify the standard ``no-slip'' boundary condition of textbook continuum hydrodynamics, which states that fluid at a solid surface has no relative velocity to it \cite{FLS}. However, this assumption successfully
describes much everyday experience. Indeed, Feynman \cite{FLS} noted in his Lectures that the no-slip condition explains
why large particles are easy to remove by blowing past a surface, but small particles are not. Similarly, readers
who wash dishes have noticed that it is difficult to remove all the soap just by running water — a dishcloth is needed for effective cleaning. Why?}''.
It is then reasonable to consider surfaces where some slip occurs but there is still some stress on the fluid \cite{WD12,Peres}. This suggests to consider (in two dimensions) the Navier-slip conditions \eqref{NS-cond}, where $\alpha=\alpha(\xx)$ is a given positive twice
continuously differentiable function defined on $\partial \Omega$, which indicates the roughness of the surface. Future research with the aim of extending the results in \cite{DNN21} to this rough-setting is planned.

\end{document}